\documentclass{article}[12pt]
\usepackage{graphicx,shuffle,tikz-cd} 
\usepackage{amsfonts}
\usepackage{mathrsfs}
\usepackage{amsmath}
\usepackage{amssymb}
\usepackage{bm}
\usepackage{feynmp-auto}
\usepackage[normalem]{ulem}
\usepackage{extarrows}
\usepackage{slashed}
\usepackage{float}
\usepackage{isodateo}
\usepackage{appendix}%%added
\usepackage{jheppub1}

\def\ma{\mathcal}
\def\ie{\begin{equation}\begin{aligned}}
\def\fe{\end{aligned}\end{equation}}

\usetikzlibrary{decorations.pathmorphing}	% For Feynman Diagrams
\usetikzlibrary{decorations.markings}
\tikzset{
	photon/.style={decorate, decoration={snake}, draw=red},
	electron/.style={draw=blue, postaction={decorate},
		decoration={markings,mark=at position .55 with {\arrow[draw=blue]{>}}}},
	gluon/.style={decorate, draw=blue,
		decoration={coil,amplitude=4pt, segment length=4pt}} ,
	vector/.style={decorate, decoration={snake}, draw},
	provector/.style={decorate, decoration={snake,amplitude=2.5pt}, draw},
	antivector/.style={decorate, decoration={snake,amplitude=-2.5pt}, draw},
	fermion/.style={draw=black, postaction={decorate},
		decoration={markings,mark=at position 1.00 with {\arrow[draw=black]{>}}}},
	fermionbar/.style={draw=black, postaction={decorate},
		decoration={markings,mark=at position .55 with {\arrow[draw=black]{<}}}},
	fermionnoarrow/.style={draw=black},
	scalar/.style={dashed,draw=black, postaction={decorate},
		decoration={markings,mark=at position .55 with {\arrow[draw=black]{>}}}},
	scalarbar/.style={dashed,draw=black, postaction={decorate},
		decoration={markings,mark=at position .55 with {\arrow[draw=black]{<}}}},
	scalarnoarrow/.style={dashed,draw=black},
	electron/.style={draw=black, postaction={decorate},
		decoration={markings,mark=at position .55 with {\arrow[draw=black]{>}}}},
	bigvector/.style={decorate, decoration={snake,amplitude=4pt}, draw},
	background/.style={dashed,draw=black, postaction={decorate},
		decoration={markings,mark=at position 1 with {\arrow[draw=black]{<>}}}},
}

\tikzstyle{block} = [draw, rectangle, 
minimum height=3em, minimum width=6em]

\title{The off-shell expansion relation of the Yang-Mills scalar theory}
\author[a]{Yi-Xiao Tao}
\author[b,c]{Konglong Wu}
\affiliation[a]{Department of Mathematical Sciences, Tsinghua University, Beijing 100084, China}
\affiliation[b]{Deutsches Elektronen-Synchrotron DESY, Notkestr. 85, 22607 Hamburg, Germany}
\affiliation[c]{School of Physics and Technology, Wuhan University, No.299 Bayi Road, Wuhan 430072, China}
\emailAdd{taoyx21@mails.tsinghua.edu.cn}
\emailAdd{konglong.wu@desy.de}

\abstract{In this work, we investigated the off-shell expansion relation of the Yang-Mills scalar theory. We explicitly showed that the single-trace Berends-Giele currents in the Yang-Mills scalar theory can be decomposed into a term expressed by a linear combination of bi-adjoint scalar Berends-Giele currents and one that vanishes under the on-shell limit. We proved that the bi-adjoint scalar currents, as well as the corresponding coefficients, can be characterized by a graphic approach that was originally studied in Einstein-Yang-Mills expansion. Furthermore, we generalized the decomposition to the multi-trace case through unifying relations and established the connection both in single-trace and multi-trace graphic descriptions. Finally, we established the relations between the Yang-Mills currents and the single-trace Yang-Mills scalar currents choosing special reference orders of the Yang-Mills graphs.}

\begin{document}

\maketitle

%----------------------------------------------
\section{Introduction}
As the leading order of the scattering amplitude, the structures of the tree-level amplitudes in diverse theories play important roles in the modern amplitude theory. Meanwhile, the relations between the tree-level amplitudes of different theories have also attracted a lot of attention, e.g. the expansion relations which will be mainly discussed later. The expansion relations of tree amplitudes construct an amazing web of different theories \cite{Chiodaroli:2017ngp,Du:2017gnh,Du:2017kpo,Fu:2017uzt,Teng:2017tbo}. For example, one can expand gravity tree amplitudes as a combination of Einstein-Yang-Mills (EYM) tree amplitudes or pure Yang-Mills (YM) amplitudes. An analog of the expansion of gravity theory is the expansion of pure YM amplitudes to the Yang-Mills scalar (YMS) amplitudes or bi-adjoint scalar (BS) amplitudes, which will be mainly studied in this paper.

Expansion relations have developed very well recently. The expansion relations can be understood by the Cachazo-He-Yuan (CHY) formalism considering the relations between the CHY integrands \cite{Cachazo:2013hca,Cachazo:2013iea,Cachazo:2014xea,Du:2017kpo,Teng:2017tbo,Stieberger:2016lng,Nandan:2016pya}. We can also use different bases to express the expansion relations so that the expansion coefficients can be given by some graphic rules \cite{Tian:2021dzf,Hou:2018bwm,Du:2019vzf}, which will also give the Bern-Carrasco-Johannson (BCJ) numerators \cite{Bern:2008qj,Bern:2010ue}. However, unlike the elegant results for amplitudes, an off-shell view of the expansion relations is still unknown. The off-shell expansion relation is a Feynman-rule version of the amplitude expansion relation, and it will help to construct the expansion relation for loop integrands. To fill this gap, we start with the simplest off-shell object: Berends-Giele (BG) current \cite{Berends:1987me}. The definition of an $n$-pt BG current is an ($n$+1)-pt tree amplitude with one leg off-shell, which allows us to construct BG currents recursively by ``gluing" vertices and the off-shell legs of some subcurrents and is easy to be generalized to different cases \cite{Armstrong:2022mfr,Tao:2023wls,Mafra:2015vca,Mafra:2010jq,Huang:2024vrb,Chattopadhyay:2024kdq}. Obviously, the recursion of BG currents is equivalent to the recursion of tree amplitudes. Moreover, the existence of the recursion allows us to prove relations about BG currents by induction. Pioneering work for the expansion relation of BG currents of YM theory can be seen in \cite{Wu:2021exa}. In that work, the authors also generalized the graphic rules (we will explain this concept later) for on-shell tree amplitudes to the BG currents. However, there are still some questions remaining. The most urgent one is the expansion relation for the YMS tree amplitudes with any number of scalar traces. In this paper, we have solved this question using some differential operators \cite{Cheung:2017ems} and the same method in \cite{Wu:2021exa}. We will also show how to reconstruct multi-trace YMS graphic rules from the single-trace ones. Moreover, we have studied how to obtain single-trace YMS results from the YM results.

This paper is organized as follows. In Section \ref{sec2}, we will review some concepts, including graphic rules, BG currents, and unifying relations. In Section \ref{sec3}, we consider the single-trace currents and prove the expansion relation for these currents by induction, using the method in \cite{Wu:2021exa}. In Section \ref{sec4}, we will show how to obtain the expansion relation of the multi-trace YMS currents from the single-trace YMS currents through the unifying relations. In this process, the emergence of the multi-trace YMS graphic rules from the single-trace ones will be done. In Section \ref{sec5}, we will demonstrate how to obtain the single-trace YMS expansion relations from the YM results. Although we failed to reconstruct the single-trace YMS graphic rules from the YM ones, we still figured out the connection between the single-trace YMS and the YM results by choosing some special reference orders.

%-------------------------------------
\section{Preliminaries}\label{sec2}

In this section, we will make a brief review on basic concepts, including  BG currents in BS and YMS, the graphic rules in YMS theory, as well as unifying relations.

%-------------------------------------
\subsection{BG recursion in BS}

The BG current  $\phi(1,...,n-1\big|\sigma_{1},...,\sigma_{n-1})$ in BS theory is defined by \cite{Mafra:2016ltu}
\begin{eqnarray} 
\phi\big(1,...,n-1\big|\sigma_{1},...,\sigma_{n-1}\big)&=&\frac{2}{s_{1...n-1}}\sum_{i=1}^{n-2}\Big[\phi\big(1,...,i\big|\sigma_{1},...,\sigma_{i}\big)\phi\big(i+1,...,n-1\big|\sigma_{i+1},...,\sigma_{n-1}\big) \nonumber \\
&&~~~~~~~~~~-\phi\big(1,...,i\big|\sigma_{n-i},...,\sigma_{n-1}\big)\phi\big(i+1,...,n-1\big|\sigma_{1},...,\sigma_{n-i}\big)\Big],\label{Eq:BerendsGieleBS}
\end{eqnarray}
where $\pmb{\sigma}=\{\sigma_{1},...,\sigma_{n-1}\}$ is a permutation of external lines $1,...,n-1$. The starting point of the recursion is
\begin{eqnarray}
 \phi(l|l')=\left\{
              \begin{array}{cc}
                1& (\,l'=l) \\
                0 & (\,l'\neq l) \\
              \end{array}\right ..
\end{eqnarray}
As a result of (\ref{Eq:BerendsGieleBS}), the BS current $\phi\big(a_1,...,a_i\big|b_1,...,b_i\big)$ has to vanish when  $\{a_1,...,a_i\}\setminus\{b_1,...,b_i\}\neq\emptyset$. The on-shell BS amplitude $A(1,...,n|\sigma_{1},...,\sigma_n)$ is then obtained by taking the following  limit
\begin{eqnarray}
A(1,2,...,n|\sigma_{1},...,\sigma_n)=\Big[s_{1...n-1}\phi\big(1,2,...,n-1\big|\sigma_{1},...,\sigma_{n-1}\big)\Bigr]\Big|_{s_{1...n-1}=k_n^2=0}.
\end{eqnarray}
The BS current (\ref{Eq:BerendsGieleBS}) satisfies many important relations which were first founded in YM theory:

\begin{itemize}
\item Reflection relation
\begin{eqnarray}
\phi\big(1,...,n-1\big|\sigma_{1},...,\sigma_{n-1}\big)=(-1)^{n}\phi\big(1,...,n-1\big|\sigma_{n-1},...,\sigma_{1}\big).\label{Eq:BSBGProperty-1}
\end{eqnarray}
\item Kleiss-Kuijf (KK) relation \cite{Kleiss:1988ne}
\begin{eqnarray}
\phi\big(1,2,...,n-1\big|\pmb{\beta},1,\pmb{\alpha}\big) =\sum_{\shuffle}(-1)^{|\pmb{\beta}|}\phi\big(1,2,...,n-1\big|1,\pmb{\alpha}\shuffle \pmb{\beta}^T\big).\label{Eq:BSBGProperty-2}
\end{eqnarray}
\item Two generalized KK relations
\begin{eqnarray}
\sum_{\shuffle}\phi\big(1,2,...,n-1\big|\pmb{\alpha}\shuffle\pmb{\beta}\big)&=&0,\label{Eq:BSBGProperty-4}\\
\sum_{\shuffle}\phi\big(1,2,...,n-1\big|\pmb{\beta}\shuffle\pmb{\gamma}^T,1,\pmb{\alpha}\big)&=&\sum_{\shuffle}(-1)^{|\pmb{\gamma}|}\phi\big(1,2,...,n-1\big|\pmb{\beta},1,\pmb{\alpha}\shuffle\pmb{\gamma}\big).\label{Eq:BSBGProperty-3}
\end{eqnarray}

\end{itemize}
In these relations $\pmb{\alpha}$, $\pmb{\beta}$ and $\pmb{\gamma}$ stand for ordered sets. The $|\pmb{\beta}|$, $|\pmb{\gamma}|$  denote the number of elements in $\pmb{\beta}$ and $\pmb{\gamma}$, while $\pmb{\beta}^T$ is the inverse permutation of $\pmb{\beta}$. The shuffling permutations $\pmb{A}\shuffle \pmb{B}$ of two ordered sets $\pmb{A}$ and $\pmb{B}$ are defined by all those permutations obtained by merging $\pmb{A}$ and $\pmb{B}$ together such that the relative order of elements in each set is preserved. The reflection relation (\ref{Eq:BSBGProperty-1}) is apparently the KK relation (\ref{Eq:BSBGProperty-2}) in the special case $\pmb{\alpha}=\emptyset$, the relation (\ref{Eq:BSBGProperty-4}) was proved by the KK relation (as pointed in \cite{Du:2011js}), while the relation (\ref{Eq:BSBGProperty-3}) can also be proven by the KK relation (\ref{Eq:BSBGProperty-2}) straightforwardly. The (color-ordered) BG currents for the YMS theory can also be constructed similarly. In this paper, we set the color order to be the natural number order for the color involving both gluons and scalars. Therefore, we will omit the label of this color order and focus on the other color order that involves only scalars.

\subsection{BG current in YMS}
The Lagrangian of YMS theory has the following expression  \cite{Bern:2019prr},
\ie
\ma{L}_{\text{YMS}}=-\frac{1}{4}F^{a}_{\mu\nu}F^{a\mu\nu}+\frac{1}{2}(D_{\mu}\phi^{A})^{a}(D^{\mu}\phi^{A})^{a}-\frac{g^2}{4}f^{abe}f^{ecd}\phi^{Aa}\phi^{Bb}\phi^{Ac}\phi^{Bd}+\frac{g\lambda}{3!} F^{ABC}f^{abc}\phi^{Aa}\phi^{Bb}\phi^{Cc}. \label{Eq:LagrangianofYMS}
\fe
It includes the gluon field strength $F^a_{\mu\nu}=\partial_{\mu}A^a_{\nu}-\partial_{\nu}A^a_{\mu}+gf^{abc}A^b_{\mu}A^c_{\nu}$, the covariant derivative $D_{\mu}\phi^{Aa}=\partial_{\mu}\phi^{Aa}+g f^{abc} A^b_{\mu}\phi^{Ac}$, and a scalar field $\phi^{Aa}$ that is charged under two gauge groups whose structure constants are $iF^{ABC}$ and $if^{abc}$ respectively. The couplings of these two gauge groups are $\lambda$ and $g$. Here we will choose $g=1$ and $\lambda=2$ for convenience\footnote{For different coupling choices, the unifying relations, which will be considered below, will get some extra constant factors. A suitable normalization will cancel this factor \cite{Cheung:2017ems}, which is equivalent to the coupling choice here.}.

The YMS BG current $J_{\text{YMS}}(\pmb{1}|\pmb{2}|\cdots|\pmb{m};\mathsf{G})$ can be calculated by the perturbiner method\footnote{In this paper, we always choose the Feynman gauge for YMS currents.} \cite{Mizera:2018jbh,Chen:2023bji}. In particular, we denote the single-trace BG current as $J_{\text{YMS}}(1,\cdots,n-1)$, in which the particles of the ordered set $\{1,2,...,n-1\}$ can be either scalar or gluons, and the last particle $``n"$ is always fixed as a scalar and in general off-shell: $k_n^2\neq0$, thus the current $J_{\text{YMS}}$ refers to a scalar. For simplicity, the color order involving only all scalars is chosen as an ordered subset of the ordered set $\{1,2,...,n-1\}$ and we will always omit this color order in our notation when considering the single-trace case. For example, for a color order involving all particles $\{1,2,3_g,4_g,5\}$, the scalar color order is $\{1,2,5\}$. 

The $J_{\text{YMS}}(1,\cdots,n-1)$ has the following expression
\begin{eqnarray}
J_{\text{YMS}}(1,\cdots,n-1)
&=&\sum_{i=1}^6 T_i, \label{Eq:YMSBGcurrent}
\end{eqnarray}
and each $T_j$ ($i=1,...6$) in above equation is given by 

%%%
\begin{eqnarray}
T_1&=&\frac{1}{s_{1,...,n-1}}\sum_{i=2}^{n}~2 J_{\text{YMS}}(1,\cdots,i-1)\, J_{\text{YMS}}(i,\cdots,n-1), \nonumber \\
T_2&=&\frac{1}{s_{1,...,n-1}}\sum_{i=2}^{n}~J_{\text{YMS}}(1,\cdots,i-1)\Bigl[J_{\text{YM}}(i,\cdots,n-1)\cdot\bigl(-2k_{1,i-1}-k_{i,n-1}\bigr)\Bigr],  \nonumber \\
T_3&=&\frac{1}{s_{1,...,n-1}}\sum_{i=2}^{n}~\Bigl[J_{\text{YM}}(1,\cdots,i-1)\cdot\bigl(2k_{i,n-1}+k_{1,i-1}\bigr)\Bigr] J_{\text{YMS}}(i,\cdots,n-1),  \nonumber \\
T_4&=&\frac{1}{s_{1,...,n-1}}\sum_{1<j<i<n}\,2\Bigl[J_{\text{YM}}(1,\cdots,i-1)\cdot J_{\text{YM}}(j,\cdots,n-1)\Bigr]\,J_{\text{YMS}}(i,\cdots,j-1), \nonumber \\
T_5&=&-\frac{1}{s_{1,...,n-1}}\sum_{1<j<i<n}\,J_{\text{YMS}}(1,\cdots,i-1)\,\Bigl[J_{\text{YM}}(i,\cdots,j-1)\cdot J_{\text{YM}}(j,\cdots,n-1)\Bigr],  \nonumber \\
T_6&=&-\frac{1}{s_{1,...,n-1}}\sum_{1<j<i<n}\Bigl[J_{\text{YM}}(1,\cdots,i-1)\cdot J_{\text{YM}}(i,\cdots,j-1)\Bigr]\,J_{\text{YMS}}(j,\cdots,n-1),\label{Eq:BGcurrentComponents}
\end{eqnarray} 
where $J_{\text{YM}}$ denotes the BG currents in pure YM \cite{Berends:1987me}, whose expansion has been summarized in Appendix~\ref{sec:ReviewOnBGCurrentYM}. The starting point of $J_{\text{YMS}}$ with an off-shell scalar leg is the same as the BS case, while in the pure YM case, the current $J_{\text{YM}}^{\mu}(l)$ represents the polarization $\epsilon_l^{\mu}$  of a gluon.  The single-trace YMS amplitude can be obtained by letting the last particle ``$n$" on-shell,
\begin{eqnarray}
A_{\text{YMS}}\left(1,2,...,n\right)
&=&\Bigl[s_{1...n-1} J_{\text{YMS}}(1,2,...,n-1)\Bigr]\Bigr|_{k_{n}^2\rightarrow0}.
\end{eqnarray}
Generally, if the last ``$n$" in $J_{\text{YMS}}(\pmb{1}|\pmb{2}|\cdots|\pmb{m};\mathsf{G})$ is always chosen as the scalar particle, the YMS amplitudes with arbitrary trace can be generated in the way
\begin{eqnarray}
A_{\text{YMS}}\left(\pmb{1}|\pmb{2}|\ldots|\pmb{m};\mathsf{G}\right)
&=&\Bigl[s_{1...n-1} J_{\text{YMS}}(\pmb{1}|\pmb{2}|\cdots|\pmb{m};\mathsf{G})\Bigr]\Bigr|_{k_{n}^2\rightarrow0}.
\end{eqnarray}

%----------------------------
\subsection{Graphic rules in YMS theory}\label{sec:YMSGraphicRules}

\begin{figure}
	\centering
    \includegraphics[width=0.9\textwidth]{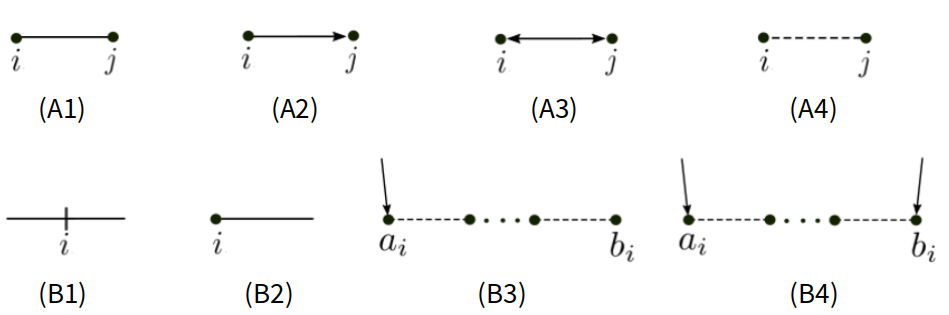}
	\caption{(A1), (A2), and (A3) represent the elements $\epsilon_i \cdot \epsilon_j$, $\epsilon_i \cdot k_j$, and $k_i \cdot k_j$, respectively. (A4) denotes two scalar particles $i$ and $j$ connected by a dashed line. (B1)--(B4) illustrate the graphical rules for constructing $\mathcal{C}^\mathcal{F}$ in Step-3: (B1) a gluon at the start of a chain contributes $\epsilon_i^\mu$; (B2) a gluon at an internal point contributes $F_i^{\mu\nu}$; (B3) a scalar trace at the start of a chain contributes $-k_{a_i}^\mu$ for all $a_i \ne b_i$; (B4) a scalar trace within a chain contributes $(k_{b_i}^\mu)(-k_{a_i}^\nu)$.}
	\label{Fig:GraphicRule}
\end{figure}

Here we briefly review a graphic approach to calculating YMS amplitudes, which is based on the recursive expansion of EYM amplitudes. The EYM amplitude can be decomposed into a basis of EYM amplitudes with fewer gravitons and finally could be expressed by a summation of pure YM amplitudes. The expansion coefficients are polynomial functions of polarizations and momentum, i.e. $(\epsilon_i\cdot\epsilon_j)$, $(\epsilon_i\cdot k_j)$ and $(k_i\cdot k_j)$. If the three types of elements are drawn by arrows and nodes (as shown in Fig.~\ref{Fig:GraphicRule} (A1)-(A3)), the expansion coefficients as well as the permutation of YM amplitudes can be described by graphs. Similarly, the YMS amplitudes $A\left(\pmb{1}|\pmb{2}|\ldots|\pmb{m};\mathsf{G}\Vert\pmb{\gamma}\right)$ also have such expansion properties. From now on we always choose $\pmb{\gamma}=\{1,2,...,n\}$ without loss of generality and ignore this label.

An arbitrary tree level YMS amplitude $A\left(\pmb{1}|\pmb{2}|\ldots|\pmb{m};\mathsf{G}\right)$ with the gluons set $\mathsf{G}$ and the scalar traces $\pmb{1}=\{1,2,\ldots, r,n\},\pmb{2},\dots,\pmb{m}$ can be generally expressed as a combination of  tree-level color-ordered BS amplitudes
\begin{eqnarray}
	A_{\text{YMS}}\left(\pmb{1}|\pmb{2}|\ldots|\pmb{m};\mathsf{G}\right)
	&=&\sum_{\mathcal{F}} (-)^{\mathcal{F}}\mathcal{C}^{\mathcal{F}}\biggl[\,\sum_{\pmb{\sigma}\in{\mathcal{F}|_1\setminus\{1,n\}}}A_{\text{YMS}}\big(1,\pmb{\sigma},n\big)\,\biggr],\label{Eq:YMSxpansion}
\end{eqnarray}
where we have summed over all possible connected tree graphs $\mathcal{F}$ which are constructed according to the graphic rules \cite{Du:2017gnh,Du:2019vzf}. The $"n"$ is fixed as the last point of the first trace $\pmb{1}$.

Fig.~\ref{Fig:GraphicRule} (A1)-(A3) are shown to represent three components $\epsilon_i\cdot\epsilon_j$, $\epsilon_i\cdot k_j$ and $k_i\cdot k_j$. The antisymmetric tensor $F_i^{\mu\nu}=\left[k_i^\mu\epsilon_i^\nu-(\mu\leftrightarrow\nu)\right]$ is drawn by a vertical line in Fig.~\ref{Fig:GraphicRule} (B1). The graph that two nodes are connected by a dashed line (in Fig.~\ref{Fig:GraphicRule} (A4)) is also introduced to denote the relative positions of scalar particles. The graphic rules in YMS amplitudes are given as follows.

\begin{itemize}
	
\item \textbf{Step-1}~~The nodes of the ordered set $\pmb{1}=\{1,...,r,n\}$ are connected with dashed lines. In particular, node $``1"$ and $``n"$ are chosen as two endpoints of this chain \footnote{In fact, arbitrary two nodes can be chosen as the endpoints.}.  The set $\mathbb{R}=\{r_1,...,r_{l+m-1}\}$ is defined with a given reference order: $r_1\succ r_2\succ...\succ r_{l+m-1}$, and it can be any permutation of the traces $\pmb{2},...,\pmb{m}$ and gluons $g_1,...,g_l$. A root set $\mathcal{R}=\{1,...,r\}$ is also defined.

\item \textbf{Step-2}~~ Define a chain with the elements in $\mathbb{R}$: $r_{i_1}\rightarrow r_{i_2}\cdots\rightarrow r_{i_k}\rightarrow j$. The starting point $r_{i_1}$ of the chain corresponds to the element with the highest priority in $\mathbb{R}$ (here it refers to $r_1$), while the internal points $r_{i_2},...,r_{i_k}$ could be any other elements in $\mathbb{R}$. The chain is attached to an arbitrary node $j$ of the root set $\mathcal{R}$. Furthermore, a new set $\mathbb{R}$ is defined as $\mathbb{R}\setminus\{r_{i_1},...,r_{i_k}\}$, and the elements of this chain have been removed from original $\mathcal{R}$. The root set is re-expressed by   $\mathcal{R}\cup\{r_1,r_{i_1},...,r_{i_k}\}$, which is the union of original root set $\mathcal{R}$ and $\{r_{i_1},...,r_{i_k}\}$. Following the procedure and constructing more chains until the set $\mathbb{R}$ becomes empty, we finally have constructed a sum of connected graphs with all particles in $\pmb{1},...,\pmb{m}$ and $\mathsf{G}$.

\item \textbf{Step-3}~~  As shown in Fig.~\ref{Fig:GraphicRule} (B2) and (B1), the gluon $g_i$ represents a polarization vector $\epsilon_i$ if it appears as the starting point, and a tensor $F_i^{\mu\nu}$ for internal point of the chains in step-2, respectively. For scalar trace $\pmb{i}$, we need to search for all possible pair $(a_i,b_i)$, for which the trace $\pmb{i}$ has the expression: $\pmb{i}=\{a_i,\pmb{\alpha},b_i,\pmb{\beta}\}$. The trace becomes $\{a_i,\pmb{\alpha}\shuffle\pmb{\beta}^T,b_i\}$ by shuffling $\pmb{\alpha}$ and $\pmb{\beta}$, which can be decomposed into a sum of chains with $a_i$ and $b_i$ as endpoints. If a trace $\pmb{i}$ appears as the starting point of any chains in step-2 (shown in Fig.~\ref{Fig:GraphicRule} (B3)), an arbitrary $b_i$ is fixed, while all possible $a_i$ of pair $(a_i,b_i)$ ($a_i\neq b_i$) are connected to second point of this chain, contributing a vector $(-k^\mu_{a_i})$ to the graph $\mathcal{C}^\mathcal{F}$. When the trace appears as the internal point of each chain in Fig.~\ref{Fig:GraphicRule} (B4), $b_i$ and $a_i$ will be connected to left- and right-hand adjacent points of the chain, respectively, contributing a term $(k_{b_i}^\mu)(-k_{a_i}^\nu)$ to the graph $\mathcal{C}^\mathcal{F}$. The ending point $j$ of each chain (in Step-2) denotes a vector $k_j^\mu$.

\item\textbf{Step-4}~~ The permutations $\pmb{\sigma}\in{\mathcal{F}|_1\setminus\{1,n\}}$ relating to the graph $\mathcal{F}$ in eq.~(\ref{Eq:YMSxpansion}) are determined by the following two steps: (i). For two adjacent nodes $a$ and $b$ satisfying that $a$ is nearer to the gluon $1$ than $b$, we have $\sigma^{-1}(a)<\sigma^{-1}(b)$ where $\sigma^{-1}(a)$ and $\sigma^{-1}(b)$ respectively denote the positions of $a$ and $b$ in $\pmb{\sigma}$.\footnote{Supposing that the position of $a$ in $\pmb{\sigma}$ is $j$, we have $a=\sigma_j\equiv{\sigma}(j)$, hence it is reasonable to define $j={\sigma}^{-1}(a)$.} (ii). For two subtree structures attached to the same node, we shuffle the corresponding permutations together such that the permutation in each subtree remains unchanged.

\item\textbf{Step-5}~~ Each trace $\pmb{i}$ contributes a $(-)^{\mathcal{F}}=(-1)^{|\pmb{\beta}|}$ for given $a_i$ and $b_i$, where $|\pmb{\beta}|$ is the number of elements in $\pmb{\beta}$ for each $\pmb{i}=\{a_i,\pmb{\alpha}\shuffle\pmb{\beta}^T,b_i\}$. 

\end{itemize}

The YMS amplitude (\ref{Eq:YMSxpansion}) can be calculated by summing over all possible graphs $\mathcal{F}$ described by the above rules. For each graph of single-trace YMS amplitudes, only gluon particles appear on the chains (of step-2), of which the graphic rule is much easier than that of pure YM amplitudes \cite{Wu:2021exa}. Moreover, since particles in $\pmb{2},...,\pmb{m}$ and $\mathsf{G}$ are not connected to $``n"$, we would be able to generalize the graphic rules by letting the last particle $``n"$ off-shell. The current defined by off-shell graphic rules (by letting $``n"$ off-shell) is shown as
%%%
\begin{eqnarray}
\tilde{J}_{\text{YMS}}\left(\pmb{1}|\pmb{2}|\ldots|\pmb{m};\mathsf{G}\right)
&=&\sum_{\mathcal{F}} (-)^{\mathcal{F}}\mathcal{C}^{\mathcal{F}}\biggl[\,\sum_{\pmb{\sigma}\in{\mathcal{F}|_1\setminus\{1,n\}}}\phi\big(1,...,n-1|1,\pmb{\sigma}\big)\,\biggr] \nonumber \\
&=&\sum_{\pmb{\sigma}^\mathcal{F}}N_{\text{YMS}}(1,\pmb{\sigma}^\mathcal{F})\,\phi(1,...,n-1|1,\pmb{\sigma}^\mathcal{F}),\label{Eq:YMSBGCurrentExpansion}
\end{eqnarray}
which is called the effective current.

%-------------------------------------------
\subsection{Unifying relations}
There exist some differential operators that can transform the amplitudes of pure YM theory into those of YMS theory, which are determined by on-shell kinematics and gauge invariance \cite{Cheung:2017ems}. In addition, these operators can also connect gravity, the non-linear sigma model, and so on. Since this work focuses mainly on the unifying relations among the YMS amplitudes (or BG currents) with different numbers of scalar traces, here we will demonstrate the relationship between the YM theory and the YMS theory. For some more studies of these differential operators, see \cite{Feng:2019tvb,Zhou:2018wvn,Zhou:2020umm,Zhou:2021kzv}.

Color-ordered YM amplitudes can be turned into YMS amplitudes through the relation:
\begin{equation}\label{eq:unifyingYMSandYM}
	A_{\rm YMS\,}=\mathcal{T}[i_{1}j_{1}]\,\mathcal{T}[i_{2}j_{2}]\,\cdots\,\mathcal{T}[i_{n}j_{n}]\cdot A_{\rm YM\,}
\end{equation}
where $\ma{T}[i_sj_s]$ with $(s=1,...,n)$ are differential operators, $i_s$ and $j_s$ refer to $i$-th and $j$-th gluon particles in YM amplitudes $A_{\rm YM\,}$, respectively. The equation above is exactly the unifying relation between YM amplitudes and YMS amplitudes.
Note that the operator $\mathcal{T}[{i_sj_s}]$ is defined as:
\begin{equation}
\mathcal{T}[{i_sj_s}]\equiv
 \partial_{\epsilon_{i_s}\epsilon_{j_s}} =\frac{\partial}{\partial\,(\epsilon_{i_s}\cdot\epsilon_{j_s})},
\end{equation}
where $\epsilon_{i_s}$ denotes the polarization vector of the $i_s$-th particle. Furthermore, the differential operator is generally shown as $\mathcal{T}[\pmb{\alpha}]$, where $\pmb{\alpha}=\{\alpha_1,...,\alpha_n\}$, and $|\alpha|\geq2$. The $\ma{T}[\pmb{\alpha}]$ can be expressed as:
\ie
\mathcal{T}[\pmb{\alpha}]=\mathcal{T}_{\alpha_{1}\,\alpha_{n}}\cdot\prod_{i=2}^{n-1}\mathcal{T}_{\alpha_{i-1}\,\alpha_{i}\,\alpha_{n}},
\fe
in which $\ma{T}_{ij}=\ma{T}[i\,j]$ and $\mathcal{T}_{ijl}=\partial_{k_{i}\epsilon_{j}}-\partial_{k_{l}\epsilon_{j}}$, and $k_{i}$ refers to the momentum of the $i$-th particle.

Note that the operators $\mathcal{T}[\pmb{\alpha}]$ are also referred to as ``trace operators", and all the elements contained in $\pmb{\alpha}$ (i.e. $\pmb{\alpha}$ would be regarded as an arbitrary permutation of these elements) are in the same trace if $\mathcal{T}[\pmb{\alpha}]$ acts on a certain amplitude. In addition, the operator $\mathcal{T}[\pmb{\alpha}]$ is invariant under the cyclic permutations of the trace $\pmb{\alpha}$.
For example, the operator $\mathcal{T}[123]$ is given by
\begin{eqnarray}
\mathcal{T}[123]
&=&\mathcal{T}_{12}\mathcal{T}_{123}
=\partial_{\epsilon_1\,\epsilon_2}\left[\partial_{k_1\,\epsilon_2} -\partial_{k_3\,\epsilon_2} \right].
\end{eqnarray}
If the operator acts on four-point YM amplitude $A_{\rm YM}(1,2,3,4)$, we immediately have a single-trace YMS amplitude with the scalar trace consisting of 1, 2, and 3:
\begin{equation}
	A_{\rm YMS}(1,2,3,4_g)=\mathcal{T}[123]\cdot A_{\rm YM}(1,2,3,4).
\end{equation} 

Finally, it is important to mention that the unifying relations also hold for BG currents under the same gauge, which can be proved by induction directly \cite{Tao:2022nqc,Chen:2023bji}. The relation will hold on the off-shell level when we choose $g=1$ and $\lambda=2$ in the YMS Lagrangian (\ref{Eq:LagrangianofYMS}). For ($n-1$)-point YMS currents with an off-shell scalar leg, we have the following relation: 
\begin{equation}
	J_{\text{YMS}}=(-1)^{\sum_{i=1}^{m}|\pmb{\alpha}_{i}|-m}\mathcal{T}[\pmb{\alpha}_{1}]\mathcal{T}[\pmb{\alpha}_{2}]\cdots\mathcal{T}[\pmb{\alpha}_{m}n](J_{\text{YM}}\cdot\epsilon_{n}). \label{Eq:UnifyingRelationBGCurrent}
\end{equation}
where $|\pmb{\alpha}_i|$ is the number of elements in $\pmb{\alpha}_i$. The relation above will be used in the following discussion.

%------------------------------------------
\section{The off-shell expansion relation of the single-trace YMS currents}\label{sec3}

The YM BG current can be decomposed into an effective current $\tilde{J}_{\text{YM}}$ (called BG current in BCJ gauge) and another two terms relating to gauge transformation \cite{Wu:2021exa}, i.e. $K_{\text{YM}}$ and $L_{\text{YM}}$, and for more details, a brief review is included in Appendix \ref{sec:ReviewOnBGCurrentYM}.

In this section, we show that the single-trace YMS BG current $J_{\text{YMS}}(1,\cdots,n-1)$ in (\ref{Eq:YMSBGcurrent}) could also be decomposed as follows:
\begin{eqnarray}
J_{\text{YMS}}(1,\cdots,n-1)
&=&\tilde{J}_{\text{YMS}}(1,\cdots,n-1)+L_{\text{YMS}}(1,\cdots,n-1), \label{Eq:DecompositionOfYMSCurrent}
\end{eqnarray}	
where $\tilde{J}_{\text{YMS}}(1,\dots,n-1)$ is the effective current in single-trace YMS that can be described by the off-shell graphic rules, while $L_{\text{YMS}}(1,\dots,n-1)$ is related to gauge transformation. Note that the particles $"1,...,n-1"$ can be either gluons or scalars. The $\tilde{J}_{\text{YMS}}(1,\dots,n-1)$  and  $L_{\text{YMS}}$ are explicitly shown as
\begin{eqnarray}
\tilde{J}_{\text{YMS}}(1,\dots,n-1)&=&\sum_{\pmb{\sigma}\in\text{Perm}(1,...,i_1-1,i_1+1,...,n-1)}\,N_{\text{YMS}}(i_1,\pmb{\sigma})\,\phi(1,\dots,n-1|\,i_1,\pmb{\sigma}), \label{Eq:SingleTraceEffectiveCurrent}
\end{eqnarray}
\begin{eqnarray}
&&~~~L_{\text{YMS}}(1,\dots,n-1) \nonumber \\
&=&\sum_{\substack{\{a_i,b_i\}\subset\{1,...,n-1\}\\\text{no scalar between $a_i$ and $b_i$}}} (-1)^{I+1}\,J_{\text{YMS}}\bigl(S_{1,a_1-1},K_{(a_1,b_1)},S_{b_1+1,a_2-1},K_{(a_2,b_2)},...,K_{(a_I,b_I)},S_{b_I+1,n-1}\bigr). \nonumber \\        \label{Eq:SingleTraceDecomposition}
\end{eqnarray}
The $i_1$ in eq.~(\ref{Eq:SingleTraceEffectiveCurrent}) denotes the first scalar in $J_{\text{YMS}}(1,\dots,n-1)$ \footnote{In this section, we denote the first and last scalar in the single-trace YMS current $J_{\text{YMS}}(1,\cdots,n-1)$ as $i_1$ and $i_2$, respectively. The reference order of the gluons is chosen as $\mathbb{R}=\{1,...,\}\setminus\pmb{1}$.}, and the numerator $N_{\text{YMS}}(i_1,\pmb{\sigma})$  can be characterized by the off-shell graphic rules. The $S_{1,a_1-1}$ refers to the sequence $(1,2,...,a_1-1)$, while $K_{(a_i,b_i)}$ is used to denote $K_{\text{YM}}(a_i,...,b_i)$ for short. The $J_{\text{YMS}}(S_{1,a_1-1},K_{(a_1,b_1)},...,K_{(a_I,b_I)},S_{b_I+1,n-1})$ stands for the single-trace BG current when $\{a_i,a_{i+1},...,b_i\}$ ($i=1,...,I$) is considered as a single external line with the polarization vector $K^{\mu}_{(a_i,b_i)}\equiv K^{\mu}_{\text{YM}}(a_i,...,b_i)$ and momentum $k^{\mu}_{a_i,b_i}$.

The BG current in YM theory satisfies an identity (\ref{Eq:IdentityYMBG-1}-\ref{Eq:IdentityYMBG-2}) if the external polarization vector $\epsilon_i$ (or arbitrary sub-current $J_{\text{YM}}(\pmb{A})$) is replaced by an off-shell momentum $k_i$ (or $k_{\pmb{A}}$). The single-trace YMS current has a similar identity with this replacement, and it can be proven following the same way as that in YM theory. Alternatively, the identity in YMS can also be derived from the unifying relations, leading to the natural vanishing of $L_{\text{YMS}}(1,\dots,n-1)$ under the on-shell limit, which will be discussed Section \ref{sec:off-shellterm}.

%--------------------------------------
\subsection{Decomposition of the single-trace YMS current}

Suppose the decomposition (\ref{Eq:DecompositionOfYMSCurrent}) is satisfied for any single-trace BG current $J_{\text{YMS}}(1,...,m)$ if $m<n-1$. Here we study the decomposition of $J_{\text{YMS}}(1,...,n-1)$.

The $T_1$ in eq.~(\ref{Eq:BGcurrentComponents}) can be decomposed in the form,
\begin{eqnarray}
T_1
&=&\frac{2}{s_{1...n-1}}\sum_{i=i_1+1}^{i_2}\,\Bigl[\tilde{J}_{\text{YMS}}(1,\cdots,i-1)\,\tilde{J}_{\text{YMS}}(i,\cdots,n-1) \nonumber \\
&&~~~~~~~~~~~~~~~~~~~~+L_{\text{YMS}}(1,\cdots,i-1)\,J_{\text{YMS}}(i,\cdots,n-1) \nonumber \\
&&~~~~~~~~~~~~~~~~~~~~+\tilde{J}_{\text{YMS}}(1,\cdots,i-1)\,L_{\text{YMS}}(i,\cdots,n-1)\Bigr].
\end{eqnarray}
 The $``i_1"$ in this case is in the sub-current $J_{\text{YMS}}(1,\cdots,i-1)$ while $``i_2"$ in $J_{\text{YMS}}(i,\cdots,n-1)$. The $T_2$ in eq.~(\ref{Eq:BGcurrentComponents}) can be rewritten as
\begin{eqnarray}
T_2
&=&\frac{1}{s_{1,...,n-1}}\sum_{i=i_2+1}^{n-1}\,\biggl\{\tilde{J}_{\text{YMS}}(1,\cdots,i-1)\,\Bigl[\tilde{J}_{\text{YM}}(i,\cdots,n-1)\cdot(-2k_{1,i-1}-k_{i,n-1})\Bigr] \nonumber \\
&&~~~~~~~~~~~~~~~~~~~~~~~~+L_{\text{YMS}}(1,\cdots,i-1)\,\Bigl[J_{\text{YM}}(i,\cdots,n-1)\cdot(-2k_{1,i-1}-k_{i,n-1})\Bigr] \nonumber \\
&&~~~~~~~~~~~~~~~~~~~~~~~~+\tilde{J}_{\text{YMS}}(1,\cdots,i-1)\,\Bigl[L_{\text{YM}}(i,\cdots,n-1)\cdot(-2k_{1,i-1}-k_{i,n-1})\Bigr] \biggr\}, \label{Eq:T2c}
\end{eqnarray}
where the decomposition (\ref{Eq:DecompositionOfYMSCurrent}) of lower-point $J_{\text{YMS}}(1,\cdots,i-1)$ has been used. The first term in the braces of above equation can be decomposed further
\begin{eqnarray}
&&\tilde{J}_{\text{YMS}}(1,\cdots,i-1)\,\Bigl[\tilde{J}_{\text{YM}}(i,\cdots,n-1)\cdot(-2k_{1,i-1})\Bigr]  \nonumber \\
&&-\tilde{J}_{\text{YMS}}(1,\cdots,i-1)\,\Bigl[K_{\text{YM}}(i,\cdots,n-1)\cdot(-k_{i,n-1})\Bigr]  \nonumber \\
&&+\tilde{J}_{\text{YMS}}(1,\cdots,i-1)\,\Bigl[\big(L_{\text{YM}}(i,\cdots,n-1) +K_{\text{YM}}(i,\cdots,n-1)\bigr)\cdot(-2k_{1,i-1}-k_{i,n-1})\Bigr] \nonumber \\
&&+\tilde{J}_{\text{YMS}}(1,\cdots,i-1)\,\Bigl[
\bigl(J_{\text{YM}}(i,\cdots,n-1)-L_{\text{YM}}(i,\cdots,n-1)\bigr)\cdot(-k_{i,n-1})\Bigr], \label{Eq:T2a-1}
\end{eqnarray}
%%%
Since $K_{\text{YM}}$  has the expression (\ref{Eq:GenK}), the second line in eq.~(\ref{Eq:T2a-1}) is equivalent to
\begin{eqnarray} \tilde{J}_{\text{YMS}}(1,\cdots,i-1)\sum_{j=i+1}^{n-1}\left(\tilde{J}_{\text{YM}}(i,\cdots,j-1)\cdot \tilde{J}_{\text{YM}}(j,\cdots,n-1)\right). \label{Eq:4pt-1}
\end{eqnarray} 
The last two lines in eq.~(\ref{Eq:T2a-1}) and last two terms in the brace of eq.~(\ref{Eq:T2c}) will contribute to $L_{\text{YMS}}(1,\cdots,n-1)$. The $T_3$ could be decomposed in a similar way, i.e.
\begin{eqnarray}
T_3
&=&\frac{1}{s_{1,...,n-1}}\sum_{i=1}^{i_1}\,\bigg\{\Bigl[\tilde{J}_{\text{YM}}(1,\cdots,i-1)\cdot(2k_{i,n-1}+k_{1,i-1})\Bigr] \tilde{J}_{\text{YMS}}(i,\cdots,n-1) \nonumber \\
&&~~~~~~~~~~~~~~~~~~+\Bigl[\tilde{J}_{\text{YM}}(1,\cdots,i-1)\cdot(2k_{i,n-1}+k_{1,i-1})\Bigr] L_{\text{YMS}}(i,\cdots,n-1) \nonumber \\
&&~~~~~~~~~~~~~~~~~~+\Bigl[\bigl(L_{\text{YM}}(1,\cdots,i-1)+K_{\text{YM}}(1,\cdots,i-1)\bigr) \cdot(2k_{i,n-1}+k_{1,i-1})\Bigr] J_{\text{YMS}}(i,\cdots,n-1)\biggr\}, \nonumber \\ \label{Eq:T3c}
\end{eqnarray}
The first term in the brace of eq.~(\ref{Eq:T3c}) can be decomposed further
\begin{eqnarray}
&&\Bigl[\tilde{J}_{\text{YM}}(1,\cdots,i-1)\cdot(2k_{i,n-1})\Bigr] \tilde{J}_{\text{YMS}}(i,\cdots,n-1) \nonumber \\
&&-\Bigl[K_{\text{YM}}(1,\cdots,i-1)\cdot(k_{1,i-1})\Bigr] \tilde{J}_{\text{YMS}}(i,\cdots,n-1) \nonumber \\
&&+\Bigl[\bigl(J_{\text{YM}}(1,\cdots,i-1)-L_{\text{YM}}(1,\cdots,i-1) \bigr)\cdot(k_{1,i-1})\Bigr] \tilde{J}_{\text{YMS}}(i,\cdots,n-1),	\label{Eq:T3a-1}
\end{eqnarray}
By using eq.~(\ref{Eq:GenK}), the second line of above equation could be simplified into
\begin{eqnarray}
-\tilde{J}_{\text{YMS}}(i,\cdots,n-1)\sum_{j=2}^{i-1}\,\bigl[\tilde{J}_{\text{YM}}(1,\cdots,j-1)\cdot \tilde{J}_{\text{YM}}(j,\cdots,i-1)\bigr]. \label{Eq:4pt-2}
\end{eqnarray}
The $T_4$, $T_5$ and $T_6$ are related to the four-point vertex, and could be decomposed in the following way
\begin{eqnarray}
T_4
&=&\frac{1}{s_{1...n-1}}\sum_{1\leq i\leq i_1,\,i_2<j\leq n-1}\, \biggl\{2\bigl[\tilde{J}_{\text{YM}}(1,\cdots,i-1)\cdot \tilde{J}_{\text{YM}}(j,\cdots,n-1)\bigr]\,\tilde{J}_{\text{YMS}}(i,\cdots,j-1)  \nonumber \\
&&~~~~~+2\Bigl[\tilde{J}_{\text{YM}}(1,\cdots,i-1)\cdot \bigl(L_{\text{YM}}(j,\cdots,n-1)+K_{\text{YM}}(j,\cdots,n-1)\bigr)\Bigr]\,\tilde{J}_{\text{YMS}}(i,\cdots,j-1)  \nonumber \\
&&~~~~~+2\Bigl[\tilde{J}_{\text{YM}}(1,\cdots,i-1)\cdot J_{\text{YM}}(j,\cdots,n-1)\Bigr]\,L_{\text{YMS}}(i,\cdots,j-1)  \nonumber \\
&&~~~~~+2\Bigl[\bigl(L_{\text{YM}}(1,\cdots,i-1)+K_{\text{YM}}(1,\cdots,i-1)\bigr)\cdot J_{\text{YM}}(j,\cdots,n-1)\Bigr]\,J_{\text{YMS}}(i,\cdots,j-1)\biggr\}, \nonumber \\
\end{eqnarray}
%%%%%%%%%%%%%%%%%%%%%
%%%%%%%%%%%%%%%%%%%%%
\begin{eqnarray}
T_5
&=&-\frac{1}{s_{1...n-1}}\sum_{ i_2<i<j\leq n-1}\,\bigg\{ \tilde{J}_{\text{YMS}}(1,\cdots,i-1)\,\bigl[\tilde{J}_{\text{YM}}(i,\cdots,j-1)\cdot \tilde{J}_{\text{YM}}(j,\cdots,n-1)\bigr] \nonumber \\
&&~~~~~~~~~+\tilde{J}_{\text{YMS}}(1,\cdots,i-1)\,\Bigl[\tilde{J}_{\text{YM}}(i,\cdots,j-1)\cdot \bigl(L_{\text{YM}}(j,\cdots,n-1)+K_{\text{YM}}(j,\cdots,n-1) \bigr)\Bigr] \nonumber \\
&&~~~~~~~~~+\tilde{J}_{\text{YMS}}(1,\cdots,i-1)\,\Bigl[\bigl(L_{\text{YM}}(i,\cdots,j-1) +K_{\text{YM}}(i,\cdots,j-1)\bigr)\cdot J_{\text{YM}}(j,\cdots,n-1)\Bigr] \nonumber \\
&&~~~~~~~~~+L_{\text{YMS}}(1,\cdots,i-1)\,\bigl[J_{\text{YM}}(i,\cdots,j-1)\cdot J_{\text{YM}}(j,\cdots,n-1)\bigr] \biggr\}, \label{Eq:DecompositionT5}
\end{eqnarray}
%%%%%%%%%%%%%%%%
%%%%%%%%%%%%%%%%
\begin{eqnarray}
T_6
&=&-\frac{1}{s_{1...n-1}}\sum_{ 1<i<j\leq i_1}\,\Bigl\{ \bigl[\tilde{J}_{\text{YM}}(1,\cdots,i-1)\cdot \tilde{J}_{\text{YM}}(i,\cdots,j-1)\bigr] \tilde{J}_{\text{YMS}}(j,\cdots,n-1) \nonumber \\
&&~~~~~~~~+\Bigl[\tilde{J}_{\text{YM}}(1,\cdots,i-1)\cdot \tilde{J}_{\text{YM}}(i,\cdots,j-1)\Bigr] L_{\text{YMS}}(j,\cdots,n-1) \nonumber \\
&&~~~~~~~~+\Bigl[\tilde{J}_{\text{YM}}(1,\cdots,i-1)\cdot \bigl(L_{\text{YM}}(i,\cdots,j-1) +K_{\text{YM}}(i,\cdots,j-1) \bigr)\Bigr] J_{\text{YMS}}(j,\cdots,n-1) \nonumber \\
&&~~~~~~~~+\Bigl[\bigl(L_{\text{YM}}(1,\cdots,i-1)+K_{\text{YM}}(1,\cdots,i-1)\bigr)\cdot J_{\text{YM}}(i,\cdots,j-1)\Bigr] J_{\text{YMS}}(j,\cdots,n-1)\biggr\}. \nonumber \\ \label{Eq:DecompositionT6}
\end{eqnarray}
The first term in the braces of eq.~(\ref{Eq:DecompositionT5}) will cancel with eq.~(\ref{Eq:4pt-1}) while that of eq.~(\ref{Eq:DecompositionT6}) is equal to eq.~(\ref{Eq:4pt-2}). 
Finally, by collecting the decomposition of each $T_i$ $(i=1,...,6)$, the effective YMS current $\tilde{J}(1,...,n-1)$ in eq.~(\ref{Eq:DecompositionOfYMSCurrent}) is a sum of $\tilde{T}_1$, $\tilde{T}_2$, $\tilde{T}_3$ and $\tilde{T}_4$
%%%%
\begin{eqnarray}
\tilde{J}(1,...,n-1)
 &=&\tilde{T}_1+\tilde{T}_2+\tilde{T}_3+\tilde{T}_4, \label{Eq:DecompositionYMSBGCurrent-2}
\end{eqnarray}
which will be proved later. Note that each $\tilde{T}_i$ ($i=1,2,3,4$) has the explicit expression
\begin{eqnarray}
\tilde{T}_1&=&\frac{2}{s_{1...n-1}}\sum_{i=i_1+1}^{i_2}\,\tilde{J}_{\text{YMS}}(1,\cdots,i-1)\,\tilde{J}_{\text{YMS}}(i,\cdots,n-1),  \label{Eq:BGDecomposition-1}  \\
\tilde{T}_2&=&\frac{1}{s_{1,...,n-1}}\sum_{i=i_2+1}^{n-1}\,\tilde{J}_{\text{YMS}}(1,\cdots,i-1)\,\Bigl[\tilde{J}_{\text{YM}}(i,\cdots,n-1)\cdot(-2k_{1,i-1})\Bigr],  \label{Eq:BGDecomposition-2}  \\
\tilde{T}_3&=&\frac{1}{s_{1,...,n-1}}\sum_{i=1}^{i_1}\,\Bigl[\tilde{J}_{\text{YM}}(1,\cdots,i-1)\cdot(2k_{i,n-1})\Bigr] \tilde{J}_{\text{YMS}}(i,\cdots,n-1), \label{Eq:BGDecomposition-3} \\
\tilde{T}_4&=&\frac{2}{s_{1...n-1}}\,\biggl\{\sum_{1\leq i\leq i_1;\,i_2<j\leq n-1}\bigl[\tilde{J}_{\text{YM}}(1,\cdots,i-1)\cdot \tilde{J}_{\text{YM}}(j,\cdots,n-1)\bigr]\,\tilde{J}_{\text{YMS}}(i,\cdots,j-1)   \nonumber \\
&&~~~~~~~~~~~~~~~~~~~~~~-\sum_{ 1<i<j\leq i_1}\,\bigl[\tilde{J}_{\text{YM}}(1,\cdots,i-1)\cdot \tilde{J}_{\text{YM}}(i,\cdots,j-1)\bigr] \tilde{J}_{\text{YMS}}(j,\cdots,n-1)\biggr\}, \nonumber \\ \label{Eq:BGDecomposition-4} 
\end{eqnarray}
while the $L_{\text{YMS}}(1,...,n-1)$ in eq.~(\ref{Eq:DecompositionOfYMSCurrent}) is given by
\begin{eqnarray}
L_{\text{YMS}}
&\sim&V_{3a}\Bigl[L_{\text{YMS}}\bigl(J_{\text{YMS}}-L_{\text{YMS}}\bigr)
+ J_{\text{YMS}}\,L_{\text{YMS}} \Bigr] +V_{3b}\Bigl[L_{\text{YMS}} J_{\text{YM}}+ \bigl(J_{\text{YMS}}-L_{\text{YMS}}\bigr)
\bigl(K_{\text{YM}}+L_{\text{YM}}\bigr) \Bigr] \nonumber \\
&&-V_{3b}\Bigl[\bigl(J_{\text{YM}}-K_{\text{YM}}-L_{\text{YM}}\bigr) L_{\text{YMS}} + \bigl(K_{\text{YM}}+L_{\text{YM}}\bigr)J_{\text{YMS}} \Bigr] \nonumber \\
&&+V_{4a}\Bigl[\bigl(J_{\text{YM}}-K_{\text{YM}}-L_{\text{YM}}\bigr)\bigl(J_{\text{YMS}}-L_{\text{YMS}}\bigr)\bigl(K_{\text{YM}}+L_{\text{YM}}\bigr) + \bigl(J_{\text{YM}}-K_{\text{YM}} \nonumber \\
&&~~~~~~~~~~-L_{\text{YM}}\bigr)L_{\text{YMS}}J_{\text{YM}} +\bigl(K_{\text{YM}}+L_{\text{YM}}\bigr)J_{\text{YMS}}\,J_{\text{YM}} \Bigr] \nonumber \\
&&+V_{4b}\Bigl[\bigl(J_{\text{YMS}}-L_{\text{YMS}}\bigr) \bigl(J_{\text{YM}}-K_{\text{YM}}-L_{\text{YM}}\bigr)\bigl(K_{\text{YM}}+L_{\text{YM}}\bigr)  +\bigl(J_{\text{YMS}}-L_{\text{YMS}}\bigr)\,\bigl(K_{\text{YM}}  \nonumber \\
&&~~~~~~~~~~+L_{\text{YM}}\bigr)J_{\text{YM}}+L_{\text{YMS}}J_{\text{YM}}\,J_{\text{YM}} \Bigr] \nonumber \\
&&-V_{4b}\Bigl[\bigl(J_{\text{YM}}-K_{\text{YM}}-L_{\text{YM}}\bigr) \bigl(J_{\text{YM}}-K_{\text{YM}}-L_{\text{YM}}\bigr)L_{\text{YMS}} +\bigl(J_{\text{YM}}-K_{\text{YM}}-L_{\text{YM}}\bigr)\,\bigl(K_{\text{YM}} \nonumber \\
&&~~~~~~~~~~+L_{\text{YM}}\bigr)J_{\text{YMS}} +\bigl(K_{\text{YM}}+L_{\text{YM}}\bigr)\,J_{\text{YM}}\,J_{\text{YMS}} \Bigr], \label{Eq:L-termSingleTraceYMS}
\end{eqnarray}
Here the vertex factors like $V_{3a}$ and $V_{4b}$ can be obtained directly by summing over the off-shell terms in $T_i$. For simplicity, we have omitted the Lorentz indices and the concrete expressions of these factors. A compact form of the off-shell terms can be given by
\ie
L_{\text{YMS}}\sim J_{\text{YMS}}\bigl(S_{1,a_1-1},K_{(a_1,b_1)},S_{b_1+1,a_2-1},K_{(a_2,b_2)},...,K_{(a_I,b_I)},S_{b_I+1,n-1}\bigr). 
\fe

%-------------------------------
\subsection{Proof of eq.~(\ref{Eq:DecompositionYMSBGCurrent-2})}

Here we show that the r.h.s. of eq.~(\ref{Eq:DecompositionYMSBGCurrent-2}) is exactly equivalent to the effective current (\ref{Eq:SingleTraceEffectiveCurrent}). Before proving this, we may introduce two useful relations for $N_{\text{YMS}}(i_1,\pmb{\sigma})$ in eq.~(\ref{Eq:SingleTraceEffectiveCurrent}). In fact, the relations between off-shell numerators constructed by graphic rules in YM theory (\cite{Wu:2021exa}) also hold in that of single-trace YMS amplitudes. 

\paragraph{Relation-1:} Given a $\pmb{\sigma}=(\pmb{\sigma}_L,\pmb{\sigma}_R)$, if $\pmb{\sigma}_L\in\text{Perm}(1,...,i_1-1,i_1+1,...,i-1)$ (with $1\leq i_1<i$) and $\pmb{\sigma}_R\in\text{Perm}(i,...,n-1)$ (with $i\leq i_2<n$), the coefficient $N_{\text{YMS}}\left(i_1,\pmb{\sigma}_L,\pmb{\sigma}_R \right)$ is equal to the multiplication of two sub-coefficients, i.e.
\begin{eqnarray}
N_{\text{YMS}}\left(i_1,\pmb{\sigma}_L,\pmb{\sigma}_R \right)
&=&N_{\text{YMS}}\left(i_1,\pmb{\sigma}_L\right)\,N_{\text{YMS}}\left(\pmb{\sigma}_R\right), \label{Eq:CoeffRelation-1}
\end{eqnarray}

\paragraph{Relation-2:} If $\pmb{\sigma}_L\in\text{Perm}(1,...,i_1-1,i_1+1,...,i-1)$ (with $1\leq i_1\leq i_2<n$) and $\pmb{\sigma}_R\in\text{Perm}(i,...,n-1)$, where $i,...,n-1$ denote the gluon particles. Thus $N_{\text{YMS}}\left(i_1,\pmb{\sigma}_L,\pmb{\sigma}_R \right)$ can be decomposed into the structure
\begin{eqnarray}
N_{\text{YMS}}\left(i_1,\pmb{\sigma}_L,\pmb{\sigma}_R \right)
&=& N_{\text{YMS}}\left(i_1,\pmb{\sigma}_L\right)\, \bigl[N_{\text{YM}}\left(\pmb{\sigma}_R\right)\cdot k_{1,i-1}\bigr], \label{Eq:CoeffRelation-2}
\end{eqnarray}
The coefficient $N_{\text{YM}}^\rho\left(\pmb{\sigma}_R\right)$ is called type-A numerator in pure YM \cite{Wu:2021exa}, and it will turn to BCJ numerator under the on-shell limit. The two equations above can be proved in the same way as eq.~(\ref{CoeffRelation-3}), which has been shown in \cite{Wu:2021exa}.

This r.h.s. of the above equation is obviously satisfied with graphic rules of the coefficient $N_{\text{YMS}}(i_1,\pmb{\sigma})$. 

Suppose eq.~(\ref{Eq:SingleTraceEffectiveCurrent}) holds for the lower-point $\tilde{J}(1,...,m)$ (with $m<n-1$), we would try to expand eqs.~(\ref{Eq:BGDecomposition-1}-\ref{Eq:BGDecomposition-3}) for the first step.

\paragraph{Expanding $\tilde{T}_1$} The $\tilde{T}_1$ can be decomposed into the following structure
\begin{eqnarray}
\tilde{T}_1
&=&\frac{2}{s_{1...n-1}}\sum_{i=i_1+1}^{i_2}\left(\sum_{\pmb{\sigma}_L} N_{\text{YMS}}(i_1,\pmb{\sigma}_L)\,\phi(1,...,i-1|i_1,\pmb{\sigma}_L)\right) \left(\sum_{\pmb{\sigma}_R} N_{\text{YMS}}(\pmb{\sigma}_R)\,\phi(i,...,n-1|\pmb{\sigma}_R)\right) \nonumber \\
&=&\frac{2}{s_{1...n-1}}\sum_{i=i_1+1}^{i_2}\sum_{\pmb{\sigma}_L}\sum_{\pmb{\sigma}_R}\Bigl[N_{\text{YMS}}(i_1,\pmb{\sigma}_L)\,N_{\text{YMS}}(\pmb{\sigma}_R)\Bigr]
\phi(1,...,i-1|i_1,\pmb{\sigma}_L)\,\phi(i,...,n-1|\pmb{\sigma}_R) \nonumber \\
&=&\frac{2}{s_{1...n-1}}\sum_{i=i_1+1}^{i_2}\sum_{\pmb{\sigma}_L}\sum_{\pmb{\sigma}_R}
N_{\text{YMS}}(i_1,\pmb{\sigma}_L,\pmb{\sigma}_R)\,\phi(1,...,i-1|i_1,\pmb{\sigma}_L)\,\phi(i,...,n-1|\pmb{\sigma}_R), \label{Eq:BGDecomposition-1a}
\end{eqnarray}
of which we have used eq.~(\ref{Eq:SingleTraceEffectiveCurrent}) in the first equality, while The second equality is maintained due to the relation (\ref{Eq:CoeffRelation-1}).

%----------------------------------
\paragraph{Expanding $\tilde{T}_2$} 

Since eq.~(\ref{Eq:BGDecomposition-2}) contains the effective currents both in YM and YMS, $T_2$ can be expanded as follows
\begin{eqnarray}
\tilde{T}_2
&=&\frac{1}{s_{1,...,n-1}}\sum_{i=i_2+1}^{n-1}\left[\sum_{\pmb{\sigma}_L} N_{\text{YMS}}(i_1,\pmb{\sigma}_L)\,\phi(1,...,i-1\,|\,i_1,\pmb{\sigma}_L)\right]
\left[\sum_{\pmb{\sigma}_R} (N_{\text{YM}}(\pmb{\sigma}_R)\cdot2k_{1,i-1})\, \phi(i,...,n-1|\pmb{\sigma}_R)\right] \nonumber \\
&=&\frac{2}{s_{1,...,n-1}}\sum_{i=i_2+1}^{n-1}\sum_{\pmb{\sigma}_L}\sum_{\pmb{\sigma}_R}\Bigl[ N_{\text{YMS}}(i_1,\pmb{\sigma}_L) (N_{\text{YM}}(\pmb{\sigma}_R)\cdot k_{1,i-1})\,\phi(1,...,i-1\,|\,i_1,\pmb{\sigma}_L)\phi(i,...,n-1|\pmb{\sigma}_R)\Bigr] \nonumber \\
&=&\frac{2}{s_{1,...,n-1}}\sum_{i=i_2+1}^{n-1}\sum_{\pmb{\sigma}_L}\sum_{\pmb{\sigma}_R}\Bigl[ N_{\text{YMS}}(i_1,\pmb{\sigma}_L,\pmb{\sigma}_R) \,\phi(1,...,i-1\,|\,i_1,\pmb{\sigma}_L)\,\phi(i,...,n-1|\pmb{\sigma}_R)\Bigr]. \nonumber \\ \label{Eq:BGDecomposition-2a}
\end{eqnarray}
where eq.~(\ref{Eq:SingleTraceEffectiveCurrent}) and eq.~(\ref{Eq:EffectiveCurrentYM}) have been used above. The second equality in (\ref{Eq:BGDecomposition-2a}) is due to the relation (\ref{Eq:CoeffRelation-2}).

\paragraph{Expanding $\tilde{T}_3$}
When $\{i_1,\pmb{\sigma}_L\}\in\text{Perm}(i,...,n-1)$ and $\pmb{\sigma}_R\in\text{Perm}(1,...,i-1)$ (which implies that the particles $``1,...,i-1"$ are gluons),  eq.~(\ref{Eq:BGDecomposition-2}) can be decomposed further
\begin{eqnarray}
\tilde{T}_3
&=&\frac{1}{s_{1,...,n-1}}\sum_{i=1}^{i_1}
\left[\sum_{\pmb{\sigma}_R}\,(N_{\text{YM}}(\pmb{\sigma}_R)\cdot2k_{i,n-1})\,\phi(1,...,i-1\,|\pmb{\sigma}_R) \right]
\left[\sum_{\pmb{\sigma}_L} N_{\text{YMS}}(i_1,\pmb{\sigma}_L)\, \phi(i,...,n-1\,|\,i_1,\pmb{\sigma}_L) \right] \nonumber \\
&=&\frac{2}{s_{1,...,n-1}}\sum_{i=1}^{i_1}\sum_{\pmb{\sigma}_L}\sum_{\pmb{\sigma}_R}
\Bigl[N_{\text{YMS}}(i_1,\pmb{\sigma}_L)\,(N_{\text{YM}}(\pmb{\sigma}_R)\cdot k_{i,n-1})\,\phi(1,...,i-1\,|\pmb{\sigma}_R)\,\phi(i,...,n-1\,|\,i_1,\pmb{\sigma}_L) \Bigr] \nonumber \\
&=&\tilde{T}_{3a}-\tilde{T}_{3b},
\end{eqnarray}
where
\begin{eqnarray}
\tilde{T}_{3a}
&=&\frac{2}{s_{1,...,n-1}}\sum_{i=1}^{i_1}\sum_{\pmb{\sigma}_L}\sum_{\pmb{\sigma}_R}
\Bigl[N_{\text{YMS}}(i_1,\pmb{\sigma}_L,\pmb{\sigma}_R)~\phi(1,...,i-1\,|\pmb{\sigma}_R)\,\phi(i,...,n-1\,|\,i_1,\pmb{\sigma}_L)\Bigr],  \label{Eq:BGDecomposition-3a} \\
\tilde{T}_{3b}
&=&\frac{2}{s_{1,...,n-1}}\sum_{i=1}^{i_1}\sum_{\pmb{\sigma}_L}\sum_{\pmb{\sigma}_R} \left(k_{j_1}\cdot k_{j_2}\right)\Bigl[\bigl(N_{\text{YMS}}(\pmb{\sigma}_R)\cdot N_{\text{YM}}(\pmb{\sigma}_{L_1})\bigr)N_{\text{YMS}}(i_1,\pmb{\sigma}_{L_2}) \nonumber \\
&&~~~~~~~~~~~~~~~~~~~~~~~~~~~~~~~~~~~~~~~~~~~~~~~\times\phi(1,...,i-1\,|\pmb{\sigma}_R)\,\phi(i,...,n-1\,|\,i_1,\pmb{\sigma}_L) \Bigr].
\end{eqnarray}
The $\tilde{T}_{3a}$ is expressed by the $(n-1)$-point $N_{\text{YMS}}(i_1,\pmb{\sigma})$.  The expression of $\tilde{T}_{3b}$ is similar to YM case (see \cite{Hou:2018bwm,Wu:2021exa}). It can be simplified further by the off-shell graph-based BCJ relation \cite{Du:2022vsw}, i.e.
\begin{eqnarray}
\tilde{T}_{3b}
&=&\frac{2}{s_{1,...,n-1}}\sum_{i=1}^{i_1}\sum_{j=i}\sum_{\pmb{\sigma}_L}\sum_{\pmb{\sigma}_R} \Bigl[\bigl(N_{\text{YMS}}(\pmb{\sigma}_R)\cdot N_{\text{YM}}(\pmb{\sigma}_{L_1})\bigr)\,N_{\text{YMS}}(i_1,\pmb{\sigma}_{L_2})\Bigr] \phi(1,...,i-1\,|\pmb{\sigma}_R) \nonumber \\
&&~~~~~~~~~~~~~~~~~~\times \Bigl[\phi(i,...,j-1\,|\,i_1,\pmb{\sigma}_{L_1})\,\phi(j,..,n-1\,|\pmb{\sigma}_{L_2}) -\bigl((i_1,\pmb{\sigma}_{L_1})\leftrightarrow\pmb{\sigma}_{L_2}\bigr)\Bigr] \nonumber \\
&=&\frac{2}{s_{1,...,n-1}}\sum_{i=1}^{i_1}\Bigl[\bigl(\tilde{J}_{\text{YM}}(1,...,i-1)\cdot\tilde{J}_{\text{YM}}(i,...,j-1)\bigr)\,\tilde{J}_{\text{YMS}}(j,...,n-1) \nonumber \\
&&~~~~~~~~~~~~~~~~~~~~-\bigl(\tilde{J}_{\text{YM}}(1,...,i-1)\cdot\tilde{J}_{\text{YM}}(j,...,n-1)\bigr)\,\tilde{J}_{\text{YMS}}(i,...,j-1)\Bigr], \label{Eq:BGDecomposition-3b}
\end{eqnarray}
which will cancel with $\tilde{T}_{4}$. The summation over eq.~(\ref{Eq:BGDecomposition-1a}), (\ref{Eq:BGDecomposition-2a}) and (\ref{Eq:BGDecomposition-3a}) are exactly equal to eq.~(\ref{Eq:SingleTraceEffectiveCurrent}).

%--------------------------------------
\section{Unifying relations for BG currents in the YMS theory}\label{sec4}

Instead of expanding the multi-trace YMS currents directly, we will show the expansion of these currents through unifying relations in this section. 

%-------------------------------------
\subsection{Unifying relations for YMS effective currents}

Here we will obtain multi-trace YMS effective currents from the single-trace ones (\ref{Eq:SingleTraceEffectiveCurrent}) through the unifying relations. Furthermore, we will also demonstrate how to obtain the graphic rules of multi-trace YMS currents from the single-trace graphic rules. 

%-------------------------------------
\subsubsection{Double trace}

Here we will prove that a double-trace current described by graphic rules can be obtained if a differential operator $\mathcal{T}[\pmb{\alpha}]$ acts on the single-trace effective current $\tilde{J}(1,...,n-1;\mathsf{G})$, where $\alpha_1,....,\alpha_m$ is the second trace generated by $\mathcal{T}[\pmb{\alpha}]$. If the double trace current is denoted by $\tilde{J}(\pmb{1}|\pmb{\alpha};\mathsf{G})$, we have the relation
\begin{eqnarray}
\tilde{J}_{\text{YMS}}(\pmb{1}|\pmb{\alpha};\mathsf{G}\setminus\pmb{\alpha})
&=&\mathcal{T}[\pmb{\alpha}]\,\tilde{J}_{\text{YMS}}(\pmb{1};\mathsf{G}) 
=\sum_{\pmb{\sigma}^\mathcal{F}}\left[\mathcal{T}[\pmb{\alpha}]\, N_{\text{YMS}}(1,\pmb{\sigma}^\mathcal{F})\right]\,\phi(1,...,n-1|1,\pmb{\sigma}^\mathcal{F}),
\end{eqnarray}
where eq.~(\ref{Eq:SingleTraceEffectiveCurrent}) has been used. Note that the $\mathsf{G}\setminus\pmb{\alpha}$ means the gluons at $\pmb{\alpha}$ turned into a scalar trace.

To show this, we need to consider two parts: the coefficient $\ma{C}^{\ma{F}}$ and the permutation set $\sigma^{\ma{F}}$. We will prove this by induction on the number of particles in the second trace.

\begin{figure}
	\centering
    \includegraphics[width=0.70\textwidth]{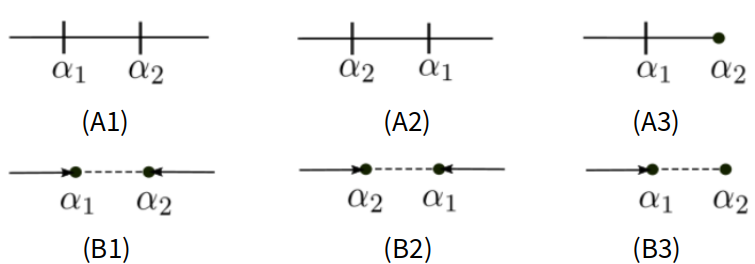}
	\caption{ The numerators $N(1,\pmb{\sigma}^\mathcal{F})$ including the sub-graphs (A1), (A2) and (A3) will be non-vanishing through differential operator $\mathcal{T}[\alpha_1,\alpha_2]$. Both $\alpha_1$ and $\alpha_2$ are internal nodes in graphs (A1) and (A2), while in $(A3)$, $\alpha_1$ and $\alpha_2$ refer to the internal and starting node of a chain. The graphs (A1), (A2) and (A3) will become (B1), (B2) and (B3) through $\mathcal{T}[\alpha_1,\alpha_2]$. }
	\label{Fig:UnifyingRelationTwoPointExample}
\end{figure}

\paragraph{Two-point differential operator $\ma{T}[{\alpha_1,\alpha_2}]$}
Let's first consider the two-point differential operator $\ma{T}[\pmb{\alpha}]$ with  $\pmb{\alpha}=\{\alpha_1,\alpha_2\}$, acting on the effective current (\ref{Eq:SingleTraceEffectiveCurrent})
\begin{eqnarray}
\mathcal{T}[\alpha_1,\alpha_2]\,\tilde{J}_{\text{YMS}}(1,...,n-1)
&=&\sum_{\pmb{\sigma}^\mathcal{F}}\left[\mathcal{T}[\alpha_1,\alpha_2]\, N_{\text{YMS}}(1,\pmb{\sigma}^\mathcal{F})\right]\,\phi(1,...,n-1|1,\pmb{\sigma}^\mathcal{F}), \label{Eq:TwoPointExample}
\end{eqnarray}
The differential operator $\mathcal{T}[\alpha_1,\alpha_2]$ is equivalent to removing the term $\epsilon_{\alpha_1}\cdot\epsilon_{\alpha_2}$ in each numerator $N_{\text{YMS}}(1,\pmb{\sigma}^{\mathcal{F}})$. The numerator $N_{\text{YMS}}(1,\pmb{\sigma}^\mathcal{F})$ of a given graph $\mathcal{F}$ will be nonzero when acted by $\mathcal{T}[\alpha_1,\alpha_2]$ only if $\mathcal{F}$ contains one of the three subgraphs Fig.~\ref{Fig:UnifyingRelationTwoPointExample} (A1) (A2) and (A3) (without loss of generality, here the reference order is chosen as $\alpha_1\prec \alpha_2$). The $\alpha_1$ and $\alpha_2$ are internal gluons of a chain in graph (A1), (A2), while in  (A3), $\alpha_1$ is an internal node adjacent to the starting point $\alpha_2$ of a chain. The graphs Fig.~\ref{Fig:UnifyingRelationTwoPointExample}  (A1), (A2) and (A3) will be changed into (B1) (B2) and (B3) through $\mathcal{T}[\alpha_1,\alpha_2]$. The $F_{\alpha_1}^{\mu\nu}F_{\alpha_2}^{\nu\rho}$ in (A1) becomes $-k_{\alpha_1}^\mu k_{\alpha_2}^\rho$ in (B1), corresponding to $(a_2,b_2)=(\alpha_1,\alpha_2)$ in Fig.~\ref{Fig:GraphicRule} (B4). Similarly, Fig.~\ref{Fig:UnifyingRelationTwoPointExample} (B2) and (B3) correspond to $(a_2,b_2)=(\alpha_2,\alpha_1)$ of Fig.~\ref{Fig:GraphicRule} (B4) and $(a_2,b_2)=(\alpha_1,\alpha_2)$ in Fig.~\ref{Fig:GraphicRule} (B3), respectively, implying that $\{\alpha_1,\alpha_2\}$ play as an scalar trace. In addition, the chains or subgraphs of a given $\mathcal{F}$ attached to Fig.~\ref{Fig:UnifyingRelationTwoPointExample} (A1), (A2) and (A3) are independent of the transformation $\mathcal{T}[\alpha_1,\alpha_2]$, for which the numerators and permutations relating to these chains will not change.

Finally, the term $\left[\mathcal{T}[\alpha_1,\alpha_2]\, N_{\text{YMS}}(1,\pmb{\sigma}^\mathcal{F})\right]$ in eq.~(\ref{Eq:TwoPointExample}) will be a numerator of a graph with scalar trace $\pmb{1}$ and $\pmb{\alpha}=\{\alpha_1,\alpha_2\}$ only  if $\mathcal{F}$ contains any subgraphs Fig.~\ref{Fig:UnifyingRelationTwoPointExample}  (A1), (A2) and (A3). The summation over all graphs $\mathcal{F}$ in eq.~(\ref{Eq:TwoPointExample}) is equivalent to that over graphs with scalar trace $\pmb{1}$ and $\pmb{\alpha}=\{\alpha_1,\alpha_2\}$, which is exactly equal to the graphic expression of $\tilde{J}_{\text{YMS}}(\pmb{1}|\pmb{\alpha};\mathsf{G}\setminus\pmb{\alpha})$.

%-------------------------------------
\paragraph{Arbitrary-point differential operator $\ma{T}[\pmb{\alpha}]$}

We first assume that for any $|\pmb{\alpha}|<m$, the double trace effective current can be obtained by acting the differential operator $\ma{T}[\pmb{\alpha}]$ on the single-trace effective current $\tilde{J}(1,...,n-1)$. Now Let's consider the operator $\ma{T}[\pmb{\alpha}]$ when $|\pmb{\alpha}|=m$, with $\pmb{\alpha}=\{\alpha_1,...,\alpha_{m-1},\alpha_{m}\}$. According to the definition, the operator can be expressed by
\begin{eqnarray}
\ma{T}[\pmb{\alpha}]
&=&\mathcal{T}[\pmb{\alpha}']\mathcal{T}_{\alpha_{m-1},\alpha_m,\alpha_1}
=\mathcal{T}_{\alpha_{m-1},\alpha_m,\alpha_1}\mathcal{T}[\pmb{\alpha}'],
\end{eqnarray}
where $\pmb{\alpha}'=\pmb{\alpha}\setminus\{\alpha_m\}$ and $\mathcal{T}_{\alpha_{m-1},\alpha_m,\alpha_1}=(\partial_{k_{\alpha_{m-1}}\epsilon_{\alpha_m}}-\partial_{k_{\alpha_{1}}\epsilon_{\alpha_m}})$, $\ma{T}[\pmb{\alpha}]$ is cyclically invariant. Based on the assumption, the double trace effective current $\tilde{J}_{\text{YMS}}(\pmb{1}|\pmb{\alpha}';\mathsf{G}\setminus\pmb{\alpha}')$ with trace $\pmb{1}$ and $\pmb{\alpha}'$ can be obtained by acting the differential operator $\mathcal{T}[\pmb{\alpha}']$ on the single-trace effective current $\tilde{J}_{\text{YMS}}(1,...,n-1)$. Here we only need to prove that the double trace effective current $\tilde{J}_{\text{YMS}}(\pmb{1}|\pmb{\alpha};\mathsf{G}\setminus\pmb{\alpha})$ is equivalent to $\mathcal{T}_{m-1,m,1}\,\tilde{J}(\pmb{1}|\pmb{\alpha}';\mathsf{G}\setminus\pmb{\alpha}')$, where
\begin{eqnarray}
\tilde{J}_{\text{YMS}}(\pmb{1}|\pmb{\alpha}';\mathsf{G}\setminus\pmb{\alpha}')
&=&
\sum_{\pmb{\sigma}^\mathcal{F}}N_{\text{YMS}}(1,\pmb{\sigma}^\mathcal{F})\,\phi(1,...,n-1|1,\pmb{\sigma}^\mathcal{F}).
\end{eqnarray}

\begin{figure}
	\centering
    \includegraphics[width=1\textwidth]{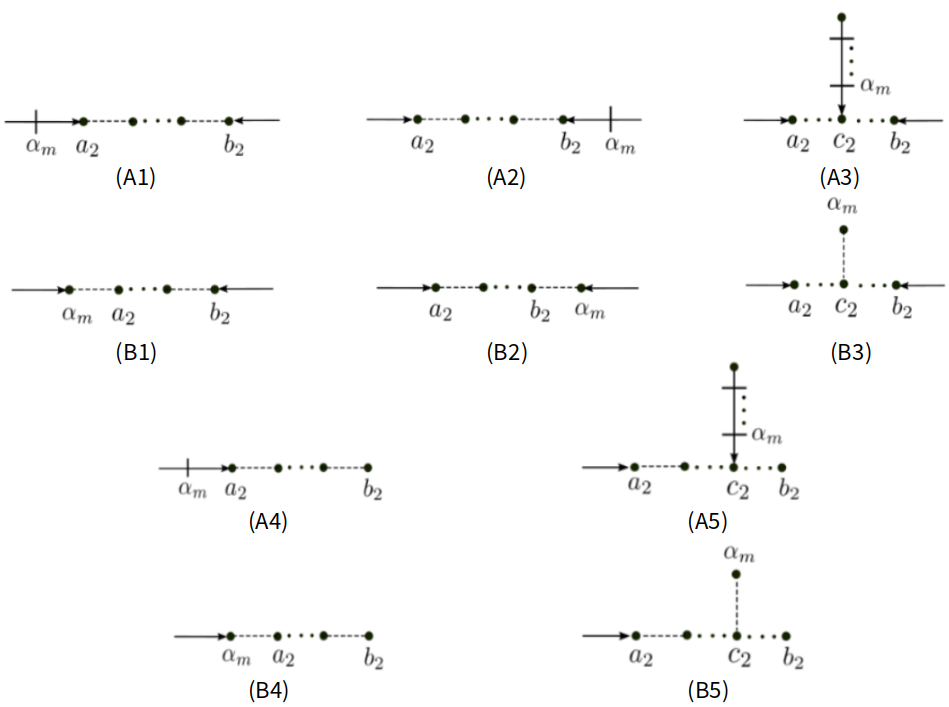}
	\caption{The numerators $N(1,\pmb{\sigma}^\mathcal{F})$ including the sub-graphs (A1), (A2) and (A3) will be non-vanishing through differential operator $\mathcal{T}_{\alpha_{m-1},\alpha_m,\alpha_1}$. The three graphs will become (B1), (B2) and (B3). The same thing happens for the starting trace case ((A4),(A5) to (B4), (B5)).}
	\label{Fig:UnifyingRelationAnyPoint}
\end{figure}

Note that when $N_{\text{YMS}}(1,\pmb{\sigma}^\mathcal{F})$ is acted by $\mathcal{T}_{\alpha_{m-1},\alpha_m,\alpha_1}$, it will be nonzero only if the graph $\mathcal{F}$ contains these subgraphs shown in Fig.~\ref{Fig:UnifyingRelationAnyPoint} (A1)-(A5). The trace $\pmb{\alpha}'$ is in the internal positions in Fig.~\ref{Fig:UnifyingRelationAnyPoint} (A1)-(A3), while starting positions in Fig.~\ref{Fig:UnifyingRelationAnyPoint}(A4)-(A5). The particle $\alpha_m$ is adjacent to $\pmb{\alpha}'$ of the same chain in (A1), (A2) and (A4), while in (A3) and (A5), a chain containing $\alpha_m$ are attached to the trace $\pmb{\alpha}'$ at $c_2\in\{\alpha_1,\alpha_{m-1}\}$. Particularly, $c_2$ is adjacent to $\alpha_m$. The five subgraphs (A1)-(A5) will turn to Fig.~\ref{Fig:UnifyingRelationAnyPoint} (B1)-(B5) under $\mathcal{T}_{\alpha_{m-1},\alpha_m,\alpha_1}$. The graphs (B1)-(B5) are discussed as follows:
\begin{itemize}
\item Fig.~\ref{Fig:UnifyingRelationAnyPoint} (A1) becomes (B1) through $\mathcal{T}_{\alpha_{m-1},\alpha_m,\alpha_1}$. The arrows in (B1) stands for the tenor $-k_{\alpha_m}^\mu k_{b_2}^\rho$. The $a_2$ in (A1) and (B1) can only take $\alpha_{1}$ and $\alpha_{m-1}$ due to the expression of $\mathcal{T}_{\alpha_{m-1},\alpha_m,\alpha_1}$.  When $(a_2,b_2)=(\alpha_1,\alpha_i)$, where $i=2,...,m-1$, the trace $\pmb{\alpha}'$ is given by $\{\alpha_1,\{\alpha_2,...,\alpha_{i-1}\}\shuffle\{\alpha_{i+1},...,\alpha_{m-1}\}^T,\alpha_i\}$ accompanied with a factor $(-1)^{m-i-1}$, thus the permutations of graph (B1) has the form
\begin{eqnarray}
(-1)(-1)^{m-i-1}\{\alpha_m,\alpha_1,\{\alpha_2,...,\alpha_{i-1}\}\shuffle\{\alpha_{i+1},...,\alpha_{m-1}\}^T,\alpha_i\}.
\end{eqnarray}
The first $``-"$ sign above comes from the operator $\mathcal{T}_{\alpha_{\alpha_{m-1}},\alpha_{m},\alpha_1}$. When $(a_2,b_2)=(\alpha_{m-1},\alpha_i)$, where $i=1,2,...,m-2$, the trace  $\pmb{\alpha}'$ is shown as $\{\alpha_{m-1},\{\alpha_1,\alpha_2,...,\alpha_{i-1}\}\shuffle\{\alpha_{i+1},...,\alpha_{m-2}\}^T,\alpha_i\}$ accompanied with a factor $(-1)^{m-i-1}$, and the permutation of graph (B2) takes the expression
\begin{eqnarray}
(-1)^{m-i}\{\alpha_m,\alpha_{m-1},\{\alpha_1,\alpha_2,...,\alpha_{i-1}\}\shuffle\{\alpha_{i+1},...,\alpha_{m-2}\}^T,\alpha_i\}.
\end{eqnarray}
 The sum of above two equations is equal to $(-1)^{m-i}\{\alpha_m,\{\alpha_1,...,\alpha_{i-1}\}\shuffle\{\alpha_{i+1},...,\alpha_{m-1}\}^T,\alpha_i\}$, which is exactly the graphic expression of scalar trace $\pmb{\alpha}$, with $\alpha_m$ and $\alpha_i$ as leftmost and rightmost endpoints (comparing with graphic rules in Section~\ref{sec:YMSGraphicRules}). 

\item Fig.~\ref{Fig:UnifyingRelationAnyPoint} (A2) becomes (B2) through $\mathcal{T}_{\alpha_{m-1},\alpha_m,\alpha_1}$. The arrows in (B2) refers to the tensor $-k_{a_2}^\mu k_{\alpha_m}^\rho$. The $b_2$ in (A2) and (B2) can only take $\alpha_{1}$ and $\alpha_{m-1}$.  When $(a_2,b_2)=(\alpha_i,\alpha_1)$, where $i=2,...,m-1$, the permutations of graph (B1) has the form
\begin{eqnarray}
(-1)(-1)^{i-2}\{\alpha_i,\{\alpha_{i+1},...,\alpha_{m-1}\}\shuffle\{\alpha_2,...,\alpha_{i-1}\}^T,\alpha_{1},\alpha_m\},
\end{eqnarray}
and if $(a_2,b_2)=(\alpha_i,\alpha_{m-1})$, where $i=1,2,...,m-2$, the permutations of graph (B1) can be
\begin{eqnarray}
(-1)^{i-1} \{\alpha_i,\{\alpha_{i+1},...,\alpha_{m-2}\}\shuffle\{\alpha_1,...,\alpha_{i-1}\}^T,\alpha_{m-1},\alpha_m\}.  
\end{eqnarray}
the sum of above two equations is equal to  $(-1)^{i-1}\{\alpha_i,\{\alpha_{i+1},...,\alpha_{m-1}\}\shuffle\{\alpha_1,...,\alpha_{i-1}\}^T,\alpha_m\}$, which is the scalar trace $\pmb{\alpha}$ (of Step-2 of graphic rules in Section~\ref{sec:YMSGraphicRules}), with $\alpha_i$ and $\alpha_{m}$ as leftmost and rightmost endpoints. 

\item Fig.~\ref{Fig:UnifyingRelationAnyPoint} (A3) becomes (B3) through $\mathcal{T}_{\alpha_{m-1},\alpha_m,\alpha_1}$. The arrows in (B3) refers to the tenor $-k_{a_2}^\mu k_{b_2}^\rho$, and the $c_2$ of this graph can be $\alpha_1$ and $\alpha_{m-1}$. If $(a_2,b_2)=(\alpha_i,\alpha_j)$ with $i<j$, the trace $\pmb{\alpha}'$ corresponds to the permutation 
\begin{eqnarray}
(-1)^{i+m-j} \{\alpha_i,\{\alpha_{i+1},...,\alpha_{j-1}\}\shuffle\{\alpha_{j+1},...,\alpha_{m-1},\alpha_1,...,\alpha_{i-1}\}^T,\alpha_j\}.
\end{eqnarray}
Correspondingly, all possible permutations of graph (B3) for $c_2=\alpha_1$ are denoted by 
\begin{eqnarray}
(-)(-1)^{i+m-j+1}\{\alpha_i,\{\alpha_{i+1},...,\alpha_{j-1}\}\shuffle\{\alpha_{j+1},...,\alpha_{m-1},\alpha_m\shuffle\{\alpha_1,...,\alpha_{i-1}\}\}^T,\alpha_j\}.
\end{eqnarray}
If $c_2=\alpha_{m-1}$, the permutation would be 
\begin{eqnarray}
(-1)^{i+m-j+1}\{\alpha_i,\{\alpha_{i+1},...,\alpha_{j-1}\}\shuffle\{\alpha_{j+1},...,\alpha_{m-1},\alpha_1,\alpha_m\shuffle\{\alpha_2,...,\alpha_{i-1}\}\}^T,\alpha_j\}.
\end{eqnarray}
The difference of above two equations is equal to 
\begin{eqnarray}
(-1)^{i+m-j+1}\{\alpha_i,\{\alpha_{i+1},...,\alpha_{j-1}\}\shuffle\{\alpha_{j+1},...,\alpha_{m-1},\alpha_m,\alpha_1,\{\alpha_2,...,\alpha_{i-1}\}\}^T,\alpha_j\},
\end{eqnarray}
which correspond to the trace $\pmb{\alpha}$ with $\alpha_i$ and $\alpha_j$ as two endpoints. The calculation above is also valid when $i>j$. 

\item Fig.~\ref{Fig:UnifyingRelationAnyPoint} (A4) becomes (B4) through $\mathcal{T}_{\alpha_{m-1},\alpha_m,\alpha_1}$. The arrows of graph (B4) represents the vector $-k_{\alpha_{m}}^\mu$. The permutations of graph (B4) is same to that of (B1), and it correspond to the trace $\pmb{\alpha}$ with $\alpha_m$ and a fixed $b_2$ as both endpoints.

\item Fig.~\ref{Fig:UnifyingRelationAnyPoint} (A5) becomes (B5) through $\mathcal{T}_{\alpha_{m-1},\alpha_m,\alpha_1}$. The arrows of graph (B5) means a vector $-k_{a_2}^\mu$. The permutations of graph (B5) is same to that of (B3), and it correspond to the trace $\pmb{\alpha}$ with $a_2$ and a fixed $b_2$ as both endpoints\footnote{Here (B4), (B5) only corresponds to the case the $\alpha_m$ not be the fixed point of the starting trace $\pmb{\alpha}$. If we want $\alpha_m$ to be the fixed point, we should change to another reference order so that $\alpha_m$ can connect to $b_2$ itself and we will obtain the graph that $\alpha_m$ is the fixed point. In this reference order (A4), (A5) will not appear and we will not obtain (B4), (B5). In other words, the arbitrariness of the fixed point of the starting trace in the new graph is ensured by the arbitrariness of the reference order of the old graph.}.
\end{itemize}

The five cases above include all possible permutations of the trace $\pmb{\alpha}$ when it is both in internal and external positions. Besides, the remaining sub-structures of a given graph connected to $\pmb{\alpha}$ will not change. It shows that the a double-trace effective current $\tilde{J}_{\text{YMS}}(\pmb{1}|\pmb{\alpha};\mathsf{G}\setminus\pmb{\alpha})$ can be obtained from single-trace ones $\tilde{J}_{\text{YMS}}(1,...,n-1)$ through the operation $\mathcal{T}[\pmb{\alpha}]$.

%-------------------------------------
\subsubsection{Multi-trace}

Since we have proved that the off-shell double trace graphic rules can be derived from single-trace version by operator $\mathcal{T}[\pmb{\alpha}]$. This will be naturally extended to $m$-trace YMS graphic rules by using a number of operator $\mathcal{T}[\pmb{P}_2]\cdots \mathcal{T}[\pmb{P}_m]$, where $\pmb{P}_i\cap\pmb{P}_j=\emptyset$ for any $i,j\in\{2,\cdots,m\}$. The corresponding current $\tilde{J}(\pmb{1}|\pmb{2}|...|\pmb{m};\mathsf{G})$ is called $m$-trace effective current in YMS,
\begin{eqnarray}
\tilde{J}_{\text{YMS}}(\pmb{1}|\pmb{2}|...|\pmb{m};\mathsf{G})
&=&\sum_{\mathcal{F}}\mathcal{T}[\pmb{P}_{2}]\cdots\mathcal{T}[\pmb{P}_{m}](N_{\text{YMS}}(1,\pmb{\sigma}^\mathcal{F}))\phi(1,..,n-1|1,\pmb{\sigma}^\mathcal{F}).
\end{eqnarray}

%%%%%%%%%%%%%%%%%%%%
\subsection{Unifying relations for off-shell terms}
Now let us turn to the off-shell terms of the YMS currents. The YMS currents with an off-shell scalar leg can be written as follows
\ie
J_{\text{YMS}}=\tilde{J}_{\text{YMS}}+L
\fe
where $J_{\text{YMS}}$ is the BG current, $\tilde{J}_{\text{YMS}}$ can be given by the graphic rules as before and actually has the same form as the corresponding YMS amplitudes, $L$ is the total off-shell terms and will vanish after taking the on-shell limit. Let $k$ be the number of on-shell scalars and $m$ be the number of scalar traces in the multi-trace current. For the single-trace current, we assume that the scalar set is $\{1,2,\cdots,r,n\}$ where $n$ is the off-shell scalar without loss of generality. Then we have the following equation where the extra minus signs come from the analysis of the BG currents from the perturbiner method \cite{Mizera:2018jbh,Tao:2022nqc}:
\ie
J_{\text{YMS}}(1,2,\cdots,r|\bold{2}|\cdots|\bold{{m}};\mathsf{G})=(-1)^{n+k-m}\sum_{\ma{F}}(-)^{\ma{F}}\ma{C}^{\ma{F}}\bigg[\sum_{\sigma^{\ma{F}}}\phi_{\text{BS}}(1,\cdots,n-1|1,\sigma^{\ma{F}})\bigg]+L_{m\text{-trace}}
\fe
and
\ie
\mathcal{T}[\pmb{P}_{2}]\cdots\mathcal{T}[\pmb{P}_{m}]J_{\text{YMS}}(1,\cdots,r,\mathsf{G})=(-1)^{k-r-m+1}J_{\text{YMS}}(1,\cdots,r|\bold{2}|\cdots|\bold{{m}},\mathsf{G}).
\fe
Applying the relation we proved for the expansion coefficient and the single-trace off-shell terms \eqref{Eq:SingleTraceDecomposition}, we have the following relation:
\ie
L_{m\text{-trace}}=(-1)^{k-r-m+1}\mathcal{T}[\pmb{P}_{2}]\cdots\mathcal{T}[\pmb{P}_{m}]L_{\text{YMS}}=\sum_{\{a_i,b_i\}\subset\{1,...,n-1\}}(-1)^{I+1}J_{\text{YMS}}(\pmb{1}|\pmb{2}|\cdots|\pmb{m};\tilde{\mathsf{G}})
\fe
where $\tilde{\mathsf{G}}$ is the division of gluons $\mathsf{G}$, and can be defined by
\begin{eqnarray}
\tilde{\mathsf{G}}=\{S'_{1,a_1-1},K_{(a_1,b_1)},S'_{b_1+1,a_2-1},K_{(a_2,b_2)},...,K_{(a_I,b_I)},S'_{b_I+1,n-1}\}.\label{multioffshell}
\end{eqnarray}
In the above expression, $S'_{1,a_1-1}$ denotes a set $\mathsf{G}\cap S_{1,a_1-1}$. The $J_{\text{YMS}}\bigl(\pmb{1}|\pmb{2}|\cdots|\pmb{m};\tilde{\mathsf{G}}\bigr)$ refers to the BG current in multi-trace YMS when $\{a_i,a_{i+1},...,b_i\}$ is considered as a single external line with the polarization vector $K_{(a_1,b_1)}$. For example, if $\mathsf{G}=\{2,4,5\}$, then $S'=\{2\}$ and $\{a_1,b_1\}=\{4,5\}$. Such expression of the off-shell terms can be obtained from the single-trace expression \eqref{Eq:SingleTraceDecomposition} directly. Note that the differential operators we considered will not affect the on-shell limit, which means that taking the on-shell limit before or after acting the differential operators on them will both lead to a zero result for off-shell terms.

The eq. \eqref{multioffshell} comes from the following statement: if we act an operator $\ma{T}[\pmb{P}]$ on the single-trace off-shell term $L_{\text{YMS}}$, then $\pmb{P}$ must be i) a subset of the legs replaced by an arbitrary $K_{\text{YM}}$ in \eqref{Eq:SingleTraceDecomposition} or ii) has no intersection with any legs in any $K_{\text{YM}}$. The former case i) will change $K_{\text{YM}}$ to $K_{\text{YMS}}$:
\begin{eqnarray}
K^{\,\rho}_{\text{YMS}}\big(1,2,...,n-1\big)=\frac{1}{s_{12\dots n-1}}\,k^{\,\rho}_{1,n-1}\,\sum_{i=1}^{n-2}\tilde{J}_{\text{YMS}}(1,...,i)\cdot \tilde{J}_{\text{YMS}}(i+1,...,n-1),\label{Eq:GenK}
\end{eqnarray}
Note that here the YMS subcurrents can have both a scalar off-shell leg and a gluon off-shell leg and can also be multi-trace currents\footnote{Although we do not consider the YMS currents with an off-shell gluon leg in this paper, the decomposition of the currents (currents as a sum of off-shell terms and effective terms which have the same expansion formalism as the amplitudes) and the unifying relations still hold in this case.}. Here we have used the fact that the effective current $\tilde{J}_{\text{YMS}}$ can be obtained by acting differential operators on $\tilde{J}_{\text{YM}}$, which will be demonstrated in the next section. The latter case ii) is just the unifying relation that changes the gluon legs to the scalar legs in the same trace. Other cases will not contribute through the following argument. Assuming that the first element of $\pmb{P}=\{P_i\}$ that intersects with a certain $K_{\text{YMS}}$ is $P_l$, then from cyclic invariant of the operator $\ma{T}[\pmb{P}]$, we can always write $\ma{T}[\pmb{P}]$ as follows:
\ie
\ma{T}[\pmb{P}]=\ma{T}[P_l,\cdots,P_{l-1}]\propto\partial_{\epsilon_{P_l}\epsilon_{P_{l-1}}}
\fe
However, as $P_{l}$ is the first element of $\pmb{P}$ intersect with that $K_{\text{YMS}}$, there will not exist any $\epsilon_{P_l}\cdot\epsilon_{P_{l-1}}$. Therefore this case will not contribute. Roughly speaking, for a given YMS current, every gluon internal line will correspond to some off-shell terms by replacing the subcurrent connecting it with corresponding $K_{\text{YMS}}$.

%---------------------------------
\section{Further study: from YM to single-trace YMS}\label{sec5}
The unifying relation also connects the YM theory and the YMS theory, and the relation is also valid off-shell. However, unlike the YMS case, we find that the full set of YMS graphic rules cannot be easily recovered from those of YM by applying differential operators. In general, acting differential operators on each graph $C^\mathcal{F}$ in pure YM commonly produces diagrams with branches attached to both endpoints, which are typically not allowed under the graphical rules of YMS. Hence the correspondence is not straightforward. Nevertheless, we still figure out some useful relations by choosing some special reference orders in the YM graphic rules. In this section, we will show how to relate the off-shell terms between the YM currents and the single-trace YMS currents and give a typical example.

%---------------------------------
\subsection{From YM expansion to single-trace YMS expansion: special reference orders}

Even though we have not figured out how to reconstruct the whole single-trace YMS graphic rules from the YM graphic rules, we can choose some special reference orders when we draw graphs so that one can obtain the single-trace YMS expansion coefficients from the YM expansion coefficients using differential operators just for this special reference order. Without loss of generality, we will demonstrate how to obtain the single-trace YMS expansion for a scalar trace $\{1,2,3,...,s\}$. Firstly, the reference order should be chosen specially so that the first and the last element of reference order is 1 and $s$ respectively. Then consider the case the $s=2$, the root of the YM graphs can only be $\{1,2\}$ and will give the correct contribution after acting $\ma{T}[12]$. Then we use the method in section \ref{sec4}, consider
\ie
\ma{T}[1,...,\alpha,\beta,s]=\ma{T}[1,...,\alpha,s](\partial_{k_{\alpha}\epsilon_{\beta}}-\partial_{k_s\epsilon_{\beta}}),
\fe
we can prove our statement by induction. Therefore, the single-trace YMS expansion coefficients from the graphic rules for a scalar trace $\{1,2,3,...,s\}$ can be obtained by acting $\ma{T}[1,2,3,...,s]$ on the YM expansion coefficients from the graphic rules.

Note that this proof is only valid for the reference order $\mathbb{R}=\{1,...,s\}$ and the particle order in $...$ is not important. If $s$ is not the last element of the reference order, this proof is not valid. As the expansion coefficients do not depend on the reference order, we exactly obtain the YMS expansion coefficients from the YM ones using differential operators. However, how to recover the whole graphic rules from the differential operators is still unknown.

%------------------------------------------
\subsection{From YM off-shell terms to single-trace YMS off-shell terms}\label{sec:off-shellterm}

Recall that the $K_{\text{YM}}^\rho$ and $L_{\text{YM}}^\rho$ are given in Appendix \ref{sec:ReviewOnBGCurrentYM}:
%%%
\begin{eqnarray}
K^{\,\rho}_{\text{YM}}\big(1,2,...,n-1\big)=\frac{1}{s_{12\dots n-1}}\,k^{\,\rho}_{1,n-1}\,\sum_{i=1}^{n-2}\tilde{J}_{\text{YM}}(1,...,i)\cdot \tilde{J}_{\text{YM}}(i+1,...,n-1),\label{Eq:GenK}
\end{eqnarray}
\begin{eqnarray}
&&~~L^{\rho}_{\text{YM}}\big(1,2,...,n-1\big)\nonumber \\
&=&\sum_{\{a_i,b_i\}\subset\{1,...,n-1\}}(-1)^{I+1}J^{\rho}_{\text{YM}}\bigl(S_{1,a_1-1},K_{(a_1,b_1)},S_{b_1+1,a_2-1},K_{(a_2,b_2)},...,K_{(a_I,b_I)},S_{b_I+1,n-1}\bigr),\label{Eq:GenL}
\end{eqnarray}
Note that only the $L_{\text{YM}}$ terms will contribute to the off-shell terms of the single-trace YMS currents through unifying relation (\ref{Eq:UnifyingRelationBGCurrent}).

The argument for the off-shell terms in this case is similar to before. However, this time we include the off-shell leg $n$ in the operator $\ma{T}[\pmb{P}_1]$ which means that $\pmb{P}_{1}$ will never be in a certain $K_{\text{YM}}$. Assuming that the first element of $\pmb{P}=\{P_i\}$ that intersects with a certain $K_{\text{YM}}$ is $P_l$, then
\ie
\ma{T}[\pmb{P}_1]\propto\ma{T}_{P_{l-1}P_ln}=(\partial_{k_{l-1}\epsilon_l}-\partial_{k_{n}\epsilon_l})
\fe
will annihilate the YM off-shell terms as we explained before. It also comes from the fact that BG currents do not include the off-shell momentum $k_n$ explicitly. Hence only the case that $\pmb{P}_1$ does not intersect with any legs in any $K_{\text{YM}}$ contributes to the single-trace YMS off-shell terms. We then have the following equation:
\begin{eqnarray}
&&~~~L_{\text{YMS}}(1,\dots,n-1) \nonumber \\
&=&\sum_{\substack{\{a_i,b_i\}\subset\{1,...,n-1\}\\\text{no scalar between $a_i$ and $b_i$}}} (-1)^{I+1}\,J_{\text{YMS}}\bigl(S_{1,a_1-1},K_{(a_1,b_1)},S_{b_1+1,a_2-1},K_{(a_2,b_2)},...,K_{(a_I,b_I)},S_{b_I+1,n-1}\bigr), 
\end{eqnarray}
which reproduces the result in section \ref{sec3}.  

%------------------------------------------
\subsection{Example: two/three-point BG currents}
In this subsection, we will demonstrate two typical examples to verify our results about the decomposition of the YMS currents.
\subsubsection{Two-point BG current in YMS} 

The YM BG current $J_{\text{YM}}(12)$ can be rewritten as the sum of $\tilde{J}^\rho_{\text{YM}}(12)$ and $K^\rho_{\text{YM}}(12)$, with
\begin{eqnarray}
s_{12}\tilde{J}^\rho_{\text{YM}}(12)
&=&2\epsilon_{1\mu}\bigl[F_2^{\mu\rho}-(\epsilon_2\cdot k_1)\eta^{\mu\rho}\bigr]
=2\bigl[\epsilon_{1\mu}F_2^{\mu\rho}-(\epsilon_2\cdot k_1) \epsilon_1^{\rho}\bigr], \nonumber \\
s_{12}K^\rho_{\text{YM}}(12)
&=&(\epsilon_1\cdot\epsilon_2)\,k^\rho_{12},
\end{eqnarray}
where $\eta^{\mu\rho}$ denotes the Minkowski metric tensor. In the above decomposition, we have used the on-shell condition $\epsilon_i\cdot k_i=k_i^2=0$ ($i=1,2$). The effective current $\tilde{J}^\rho_{\text{YM}}(12)$ will become $J_{\text{YMS}}(1,2_g)=(\epsilon_2\cdot k_1)\phi(12|12)$ under the operator $\mathcal{T}[23]$, while $K^\rho_{\text{YM}}(12)$ turns to be 0. 

%--------------------------------------------
\subsubsection{Three-point BG current in YMS}
As a typical example of our statement, consider the following process
\ie
J_{\text{YM}}(123)\cdot\epsilon_4\stackrel{\ma{T}[14]}{\longrightarrow} J_{\text{YMS}}(1_s,2_g,3_g)\stackrel{\ma{T}[23]}{\longrightarrow} J_{\text{YMS}}(1_s|2_s,3_s),
\fe
For the expansion coefficients from YM to single-trace YMS theory, see appendix \ref{sec:ReviewOnBGCurrentYM}. We will demonstrate that the off-shell terms of the above double-trace current can be obtained from the YM ones. For the 3-pt YM currents $J_{\text{YM}}^{\rho}(123)$, the off-shell terms (\ref{Eq:GenK}) and (\ref{Eq:GenL}) are 
\begin{eqnarray}
K^{\rho}_{\text{YM}}(1,2,3)&=&\frac{1}{s_{123}}\Big[\tilde{J}_{\text{YM}}(1,2)\cdot\epsilon_3+\epsilon_1\cdot\tilde{J}_{\text{YM}}(2,3)\Big]k_{1,3}^{\rho} \\
L^{\rho}_{\text{YM}}(1,2,3)&=&J_{\text{YM}}^{\rho}(K_{\text{YM}}(1,2),3)+J^{\rho}(1,K_{\text{YM}}(2,3)),
\end{eqnarray}
Then the only term that contributes to the single-trace off-shell terms is
\ie
J_{\text{YM}}^{\rho}(1,K_{\text{YM}}(2,3))
\fe
since other terms will not give a $\epsilon_1\cdot\epsilon_4$ after contracting with $\epsilon_4$. Then the single-trace off-shell terms should be
\ie
\ma{T}[14]J_{\text{YM}}(1,K(2,3))\cdot\epsilon_4=-\frac{1}{s_{23}s_{123}}(\epsilon_2\cdot\epsilon_3)\left[k_{2,3}\cdot(2k_1+k_{2,3})\right]=-\frac{\epsilon_2\cdot\epsilon_3}{s_{23}}
\fe
which is exactly the correct answer from direct calculations. The double-trace off-shell term should be
\ie
-\ma{T}[23](-\frac{\epsilon_2\cdot\epsilon_3}{s_{23}})=\frac{1}{s_{23}}
\fe
A direct calculation shows that
\ie
s_{123}J_{\text{YMS}}(1|2,3)
	=1-2\frac{k_{1}\cdot(k_{3}-k_{2})}{s_{23}}=-\frac{4k_1\cdot k_3}{s_{23}}+\frac{s_{123}}{s_{23}}\rightarrow L_{\text{YMS}}=\frac{1}{s_{23}}
\fe
which matches with the results from differential operators.

%--------------------------------------------
\section{Conclusions}
In this paper, we derived the expansion relations for any-trace YMS currents and wrote down the decomposition of the YMS currents explicitly. We first figured out the expansion relations for the single-trace current using a method similar to \cite{Wu:2021exa}. Then we pointed out how to obtain the multi-trace graphic rules by acting some differential operators on the single-trace graphic rules for amplitudes. Such differential operators, known as the unifying relations, are also valid in the off-shell case for the YM and the YMS theory \cite{Tao:2022nqc}. A YMS current with an off-shell scalar leg can always be decomposed into two parts: a part with the same expression as the corresponding amplitudes and a part that will vanish after taking the on-shell limit which is called ``the off-shell terms". Hence we can use such differential operators to obtain the expansion relations for multi-trace from the single-trace ones. Finally, we connected the YM current expansion and the single-trace YMS current expansion by choosing some special reference orders and reproduced the expression of the off-shell terms for the single-trace YM currents. In conclusion, we found that the graphic rules and the unifying relations allowed us to figure out all terms in the expansion relations explicitly for any-trace YMS currents and such relations can also be generalized to the YM currents case.

There are still some related problems that deserve further study. We list some of them here:
\begin{enumerate}

\item In this work, we only consider the expansion relations for BG currents which have only one off-shell leg. What about the case that all legs are off-shell? This question is equivalent to finding out the expansion relation for the Feynman rules.

\item Can we figure out the expansion relations for 1-loop integrands using our method? This question is equivalent to finding out the possible 1-loop BCJ numerator from an off-shell way. A possible way to reach this is by finding out the expansion relation for currents with 2 legs off-shell and then using the sewing procedure \cite{Gomez:2022dzk} to generate a loop. There is also an alternative way to consider the loop effect \cite{Lee:2022aiu}. Note that there are also some works about the 1-loop BCJ numerator from the on-shell methods \cite{Edison:2022jln,Dong:2023stt,Xie:2024pro}.

\item A harder direction is to do the same things on the BG currents for extended gravity \cite{Tao:2023yxy,Cho:2021nim,Cho:2022faq}, and then consider the second question above. This question is more meaningful since we are more interested in the gravity theory. Note that the unifying relation is only valid for amplitudes in this case, which means that the method we used in this paper cannot be generalized to the gravity case.
\end{enumerate}

%------------------------------------
\section*{Acknowledgement}
We would like to thank Yi-Jian Du for the valuable discussion. YT is partly supported by the National Key R\&D Program of China (NO. 2020YFA0713000). KW is supported by the Helmholtz-OCPC International Postdoctoral Exchange Fellowship Program. 

%------------------------------------
\appendix

%------------------------------------
\section{Review on the decomposition of BG current in YM}\label{sec:ReviewOnBGCurrentYM}

Here we briefly review on some useful results relating to YM BG current in \cite{Wu:2021exa}. The YM BG current has the form
\begin{eqnarray}
J^\mu_{\text{YM}}(1,...,n-1) 
&=&\frac{1}{s_{1...n-1}}\Bigl[~\sum_{i=2}^{n-1}\,V_{3}^{\mu\nu\rho}\,J_{\text{YM},\nu}(1,...,i-1)\,J_{\text{YM},\rho}(i,...,n-1) \nonumber \\
&&~~~~~~~~+\sum_{1<i<j<n-1}\,V_{4}^{\mu\nu\rho\gamma}J_{\text{YM},\nu}(1,...,i-1)J_{\text{YM},\rho}(i,...,j-1)J_{\text{YM},\gamma}(j,...,n-1) \Bigr], \nonumber \\
\end{eqnarray}
where $V_{3}^{\mu\nu\rho}$ and $V_{4}^{\mu\nu\rho\gamma}$ are the three-point and four-point vertex in Yang-Mills field. It has shown in \cite{Wu:2021exa} that, the BG current in the Feynman gauge was decomposed as follows
\begin{eqnarray}
J^\rho_{\text{YM}}(1,...,n-1)
=\tilde{J}^\rho_{\text{YM}}(1,...,n-1)+K^\rho_{\text{YM}}(1,...,n-1)+L^\rho_{\text{YM}}(1,...,n-1).
\end{eqnarray}
where $\tilde{J}_{\text{YM}}(1,...,n-1)$ is called the 	effective current in YM. The $K^\rho_{\text{YM}}(1,...,n-1)$ and $L^\rho_{\text{YM}}(1,...,n-1)$ are related to gauge transformation, which will vanish under the on-shell limit. The effective current is exactly the BG current in BCJ gauge \cite{Mafra:2015vca}, and it can be expressed by a sum of BS currents
\begin{eqnarray}
\tilde{J}^\rho_{\text{YM}}(1,...,n-1)
&=&\sum_{\pmb{\sigma}\in\mathcal{P}(2,...,n-1)} N^\rho_{\text{YM}}(1,\pmb{\sigma})\,\phi(1,...,n-1|\,1,\pmb{\sigma}), \label{Eq:EffectiveCurrentYM}
\end{eqnarray}
in which the coefficients $N^\rho_{\text{YM}}(1,\pmb{\sigma})$ are the off-shell BCJ numerators. Both the permutations $\pmb{\sigma}$ and $N^\rho_{\text{YM}}(1,\pmb{\sigma})$ can be characterized by graphic rules in YM. The $K_{\text{YM}}^\rho$ and $L_{\text{YM}}^\rho$ are given by
%%%
\begin{eqnarray}
K^{\,\rho}_{\text{YM}}\big(1,2,...,n-1\big)=\frac{1}{s_{12\dots n-1}}\,k^{\,\rho}_{1,n-1}\,\sum_{i=1}^{n-2}\tilde{J}_{\text{YM}}(1,...,i)\cdot \tilde{J}_{\text{YM}}(i+1,...,n-1),\label{Eq:GenK}
\end{eqnarray}
\begin{eqnarray}
&&~~L^{\rho}_{\text{YM}}\big(1,2,...,n-1\big)\nonumber \\
&=&\sum_{\{a_i,b_i\}\subset\{1,...,n-1\}}(-1)^{I+1}J^{\rho}_{\text{YM}}\bigl(S_{1,a_1-1},K_{(a_1,b_1)},S_{b_1+1,a_2-1},K_{(a_2,b_2)},...,K_{(a_I,b_I)},S_{b_I+1,n-1}\bigr),\label{Eq:GenL}
\end{eqnarray}
where $S_{1,a_1-1}$ denotes the sequence $1,2,...,a_1-1$, while $K_{(a_i,b_i)}$ refers to $K_{\text{YM}}(a_i,a_i+1...,b_i)$. The $J^{\rho}_{\text{YM}}(S_{1,a_1-1},K_{(a_1,b_1)},...,K_{(a_I,b_I)},S_{b_I+1,n-1})$ represents the BG current when $\{a_i,a_{i+1},...,b_i\}$ ($i=1,...,I$) is considered as a single external line with the polarization vector $K^{\mu}_{(a_i,b_i)}\equiv K^{\mu}_{\text{YM}}(a_i,...,b_i)$ and momentum $k^{\mu}_{a_i,b_i}$. The currents $J_{\text{YM}}$ in $L_{\text{YM}}$ will vanish under on-shell limit, i.e.
\begin{eqnarray}
&&k_{1,n-1}\cdot J_{\text{YM}}(1,...,a_1-1,k_{a_1,b_1},b_1+1,...,a_2-1,k_{a_2,b_2},...,k_{a_I,b_I},b_I+1,...,n-1)=0, \label{Eq:IdentityYMBG-1}\\
&&\epsilon_n\cdot J_{\text{YM}}(1,...,a_1-1,k_{a_1,b_1},b_1+1,...,a_2-1,k_{a_2,b_2},...,k_{a_I,b_I},b_I+1,...,n-1)=0, \label{Eq:IdentityYMBG-2}
\end{eqnarray}
which leads to the vanishing of $K^{\rho}_{\text{YM}}\big(1,2,...,n-1\big)$ and $L^{\rho}_{\text{YM}}\big(1,2,...,n-1\big)$.

The graphic rules in YM are a little different from that of single-trace YMS. The dashed line of the scalar trace in step-1 of Section \ref{sec:YMSGraphicRules} is replaced by a chain $\bigl[\epsilon_1\cdot F_{j_1}\cdots F_{j_s}\cdot\epsilon_n\bigr]$, which is from the root set $\mathcal{R}=\{1,j_1,j_2,...,j_s\}$. The $j_1,...,j_s$ are chosen arbitrarily from $\{2,...,n-1\}$. In addition, the reference order is fixed by the set $\mathbb{R}=\{1,...,n-1\}$, and $\mathbb{R}\backslash\mathcal{R}$ is defined to construct the chain towards $\mathcal{R}$. The $``1"$ and $``n"$ are fixed as two endpoints of a given graph, and by letting $``n"$ off-shell, the graphic rules can be naturally generalized into off-shell. 

Three types of off-shell numerators, including type-A/B/C numerators, have been introduced. The type-A/B numerators correspond directly to $N_{\text{YM}}^\rho(1,\pmb{\sigma})$ in the effective current. Each $N_{C}^\rho$ is a tensor that keeps two endpoints off-shell, while the branches attached satisfy the graphic rules. If $\pmb{\sigma}_L\in\text{Perm}(2,...,i-1)$ and $\pmb{\sigma}_R\in\text{Perm}(i,...,n-1)$, these numerators are connected through the relation
\begin{eqnarray}
N_{A}^\rho(1,\pmb{\sigma}_L,\pmb{\sigma}_R)=\bigl[N_A(1,\pmb{\sigma}_L) N_C(\pmb{\sigma}_R)-N_A^\mu(1,\pmb{\sigma}_L)\bigl(N_B(\pmb{\sigma}_R)\cdot k_{1,i-1}\bigr)\bigr]^\rho, \label{CoeffRelation-3}
\end{eqnarray}
which has been proved in \cite{Wu:2021exa}. The relation above can be extended to the YMS case, i.e. i) the relation (\ref{Eq:CoeffRelation-1}) is similar to  $\bigl[N_A^\mu(1,\pmb{\sigma}_L)N_C^\mu(1,\pmb{\sigma}_L)\bigr]^\rho$ when the last scalar $``i_2"$ in Section \ref{sec3} is included in $\pmb{\sigma}_R$, and ii)  the relation (\ref{Eq:CoeffRelation-2}) is similar to $\bigl[N_A^\mu(1,\pmb{\sigma}_L)\bigl(N_B(\pmb{\sigma}_R)\cdot k_{1,i-1}\bigr)\bigr]^\rho$ if $``i_2"$ is in $\pmb{\sigma}_L$ (in which $\pmb{\sigma}_R$ refer to pure gluons).  Eqs.~(\ref{Eq:CoeffRelation-1}-\ref{Eq:CoeffRelation-2}) can be proved in the same way as eq.~(\ref{CoeffRelation-3}), which will not be illustrated in the present work.

\bibliographystyle{JHEP}
\bibliography{YMS}
\end{document}